\documentclass[structabstract]{aa}  
\usepackage{natbib}
\bibpunct{(}{)}{;}{a}{}{,} 
\usepackage{graphicx}
\usepackage{txfonts}
\usepackage{lscape}
\newcommand{\hst}{{\sl HST}}
\newcommand{\hubble}{{\sl Hubble}}
\newcommand{\rosat}{{\sl ROSAT}}
\newcommand{\chandra}{{\sl Chandra}}
\newcommand{\xmm}{{\sl XMM-Newton}}

\def\nh{\hbox{$N_{\rm H}$}}

\def\hei{He\,{\sc i}}
\def\hii{H\,{\sc ii}}
\def\ha{H$\alpha$}
\def\sii{S\,{\sc ii}}
\def\oiii{O\,{\sc iii}}
\def\logfxfo{log($f_{\rm X}/f_{\rm opt}$)}

\begin{document}
\title{Supernova remnants and candidates detected in the \xmm\ M\,31 large
survey\thanks{Based on observations obtained with \xmm, an ESA science mission with instruments and contributions directly funded by ESA Member States and NASA.}}

\subtitle{}

\titlerunning{X-ray SNRs in M\,31}

\author{Manami Sasaki
  \inst{1}
  \and
  Wolfgang Pietsch\inst{2} 
  \and
  Frank Haberl\inst{2} 
  \and
  Despina Hatzidimitriou\inst{3}
  \and
  Holger Stiele\inst{4}
  \and 
          Benjamin Williams\inst{5}
          \and 
          Albert Kong\inst{6}
          \and 
          Ulrich Kolb\inst{7}
          }

   \institute{Institut f\"ur Astronomie und Astrophysik, 
              Universit\"at T\"ubingen,
              Sand 1, 
              D-72076 T\"ubingen, Germany,
              \email{sasaki@astro.uni-tuebingen.de}
         \and
              Max-Planck-Institut f\"ur extraterrestrische Physik, 
              Giessenbachstra{\ss}e,
              D-85748 Garching, Germany
         \and
              Department of Astrophysics, Astronomy and Mechanics, 
              Faculty of Physics, 
              University of Athens, 
              Panepistimiopolis, 
              15784 Zografos, 
              Athens, Greece
         \and
              INAF-Osservatorio Astronomico di Brera, 
              Via E. Bianchi 46, 
              I-23807 Merate (LC), Italy
         \and
              Department of Astronomy, 
              Box 351580, 
              University of Washington, 
              Seattle, WA, 98195, USA
         \and
              Institute of Astronomy and Department of Physics, 
              National Tsing Hua University, 
              Hsinchu 30013, Taiwan
         \and
              Dept of Physical Sciences, 
              The Open University, Walton Hall, 
              Milton Keynes,
              MK7 6AA, UK
             }

   \date{Received February 13, 2012; accepted June 20, 2012.}

 
  \abstract
   {We present the analysis of supernova remnants 
(SNRs) and candidates
in M\,31 identified in the \xmm\ large programme survey of M\,31.
SNRs are among the bright X-ray sources in a galaxy. They are
good indicators of recent star formation activities of a galaxy
and of the interstellar environment in which they evolve.}
   {By combining the X-ray data of sources in M\,31
with optical data as well as with optical and radio catalogues, 
we aim to compile a complete, revised list of
SNRs emitting X-rays in M\,31 detected with \xmm, study their luminosity
and spatial distribution, and understand the X-ray spectrum
of the brightest SNRs. }
   {We analysed the X-ray spectra of the twelve brightest SNRs and 
candidates using \xmm\ data. The four brightest sources allowed us 
to perform a more detailed
spectral analysis and the comparison of different models to describe their
spectrum. For all M\,31 large programme sources we searched for optical 
counterparts on the \ha, [\sii], and
[\oiii] images of the Local Group Galaxy Survey.}
   {We confirm 21 X-ray sources as counterparts of known SNRs. 
In addition, we identify five new X-ray sources as X-ray and optically 
emitting SNRs. 
Seventeen sources are no longer considered as SNR candidates.
We have thus created a list of 26 X-ray SNRs and 20 candidates in M\,31 
based on their X-ray, optical, and radio emission,
which is the most recent complete list of X-ray SNRs in M\,31.
The brightest
SNRs have X-ray luminosities of up to $8 \times 10^{36}$~erg~s$^{-1}$ in 
the 0.35 -- 2.0~keV band. 
} 
   {}


   \maketitle
%

\section{Introduction}

The Andromeda galaxy (M\,31) is the largest galaxy in the Local
Group and the nearest spiral galaxy to the Milky Way. Its size and 
mass are comparable to those of our Galaxy. Therefore, this archetypical
spiral galaxy provides us a unique opportunity to study and understand
the nature and the evolution of a galaxy like our own.
Various authors have studied the star formation history in different
regions of M\,31 using observations with both the \hubble\ Space
Telescope (\hst) and large ground-based telescopes 
(e.g., the Local Group Galaxy Survey [LGGS] performed at the 
Kitt Peak National Observatory [KPNO] and the Cerro Tololo
Inter-American Observatory, 
Williams \citeyear{2003AJ....126.1312W}; 
Massey et al.\ \citeyear{2006AJ....131.2478M}).
Deep {\sl HST} photometry
has shown that the mean age of the disk of M\,31 is $\sim6 - 8$~Gyr 
\citep{2006ApJ...652..323B} and that
the average metallicity is [Fe/H] $\approx$ --0.6~dex \citep{2003A&A...405..867B}.
\citet{2003AJ....126.1312W}
measured a mean star formation rate of about 1~$M_{\sun}$~yr$^{-1}$
in the full disk of M\,31. \citet{2010A&A...517A..77T} have recently
studied the dust distribution and computed the de-reddened \ha\ distribution 
in the disk of M\,31. They derived a star formation rate of
0.27~$M_{\sun}$~yr$^{-1}$ for the radial range of $6 < R < 17$~kpc
with an increase to about twice the mean value at around $R = 10$~kpc.
Although the current star formation rate in M\,31 is lower than that
of the Milky Way, M\,31 seems to have undergone more active star formation
periods.
In addition to the well-known dust ring at a radius of $\sim$10~kpc 
\citep{1984A&AS...55..179B,1993ApJ...418..730D}
with enhanced star formation, 
\citet{2006Natur.443..832B} found an inner dust ring with 
a radius of 1 -- 1.5~kpc, which has apparently been created in
an encounter with a companion galaxy, most likely with M\,32.

First observations of individual sources in M\,31 in X-rays were
performed with the {\sl Einstein} Observatory
\citep{1979ApJ...230..540G} in the energy band
of 0.2 -- 4.5~keV and yielded the first catalogues of X-ray sources in the 
field of M\,31 \citep{1979ApJ...234L..45V,1991ApJ...382...82T}.
In the 1990s, {\sl ROSAT} \citep{1982AdSpR...2..241T} observed
M\,31 in the 0.1 -- 2.4~keV band and revealed a total of 560 sources in the
field of M\,31 (Supper et al.\ \citeyear{1997A&A...317..328S}, hereafter SHP97; 
Supper et al.\ \citeyear{2001A&A...373...63S}, hereafter SHL01). 
The next generation X-ray satellites \chandra\ X-ray Observatory 
\citep{2002PASP..114....1W} 
and X-ray Multi-Mirror Mission \citep[\xmm,][]{2001A&A...365L...1J},
which were both launched in 1999,
have significantly improved spatial and spectral resolutions with respect to 
the prior X-ray telescopes. They have also performed several observations 
of M\,31 and allowed both to obtain a comprehensive list of X-ray sources
and to study individual sources
\citep[e.g.,][]{2001A&A...378..800O,2002ApJ...577..738K,
2002ApJ...578..114K,2005A&A...434..483P,2008A&A...480..599S,
2008ApJ...689.1215B}.
The entire galaxy M\,31 was observed by \xmm\ in a large programme (LP) 
between June 2006 and February 2008 with the European Photon Imaging Cameras 
\citep[EPICs,][]{2001A&A...365L..18S,2001A&A...365L..27T} as
prime instruments. The \xmm\ source catalogue with 1897 sources has been 
published by Stiele et al.\ (\citeyear{2011A&A...534A..55S}, hereafter SPH11). 

Supernova remnants are the aftermath of stellar explosions releasing a large 
amount of energy in galaxies. The spherically expanding blast wave shock 
produces a cavity in the interstellar medium with a very low-density, 
high-temperature interior, which predominantly emits soft X-ray radiation. 
In addition, relativistic electrons and heavier particles in SNRs emit 
synchrotron emission which can be detected in radio and in some cases
also in X-rays. After a few thousand years, the SNR becomes radiative, 
i.e., the radiative losses
in SNR shocks expanding into the ambient ISM become non-negligible 
and the shell emits energy as UV and optical line emission.
If a neutron star is created in the supernova explosion a pulsar and/or a 
pulsar wind nebula (PWN) can be found inside the SNR, in which particles 
are accelerated in the strong magnetic field of the neutron star and thus 
non-thermal emission is produced. 
SNRs in M\,31 were mainly detected in the optical
\citep[e.g.,][]{1980A&AS...40...67D,1981AJ.....86..989D,1981ApJ...247..879B}
and in combined optical and radio studies \citep{1993A&AS...98..327B}.
The X-ray survey performed with {\sl ROSAT}
led to the detection of 16 X-ray SNRs \citep{2001A&A...373...63S}, while
21 were detected and identified with \xmm\ \citep{2005A&A...434..483P}. 
\citet{2002ApJ...580L.125K} presented the first
resolved X-ray image of an SNR in M\,31 taken with
\chandra, while \citet{2003ApJ...590L..21K} and 
\citet{2004ApJ...615..720W} reported on the discovery of 
new SNRs in M\,31 based on \chandra\ data. 

In this paper, we present the study of all X-ray sources that have been 
suggested to be SNRs or candidates of SNRs in M\,31 based on a complete
survey of M\,31 performed within the framework of the LP of \xmm\ (SPH11). 
Through a detailed study of each SNR and candidate detected by \xmm, 
we have obtained an improved sample of SNRs in M\,31 consisting of
X-ray or optically confirmed SNRs as well as bona-fide candidates.
We have performed a spectral analysis of the bright SNRs and candidates in the 
catalogue of SPH11. Using optical data of the LGGS, we have searched 
for optical counterparts of the X-ray SNRs and candidates showing optical 
\ha, [\sii], and [\oiii] emission from the radiative shock. 
We newly calibrated the LGGS data to obtain optical fluxes and computed
the [\sii]/\ha\ flux ratio, which is an indicator of SNR emission in the
optical.
This study has 
also allowed us to identify sources that are most probably no SNRs. 
In addition, we have performed statistical studies using the revised list
of SNRs and candidates in M\,31 detected with \xmm.

\section{Data}

\subsection{X-ray data}\label{xdata}

\begin{table*}
\caption{
\label{addcandlist}
Additional X-ray sources in the \xmm\ M\,31 survey
catalogue by \citet{2011A&A...534A..55S} analysed in this work.
}
\centering
\begin{tabular}{rccccrcccc} 
\hline\hline\
[SPH11] & RA (2000.0) & Dec (2000.0) & Pos.\ error & Rate & $ML$\tablefootmark{a} & $HR_1$\tablefootmark{b} & $HR_2$\tablefootmark{b} & $HR_3$\tablefootmark{b} & $HR_4$\tablefootmark{b} \\
\multicolumn{1}{c}{ID} &  &  & [arcsec] & [cts/s] & & & & \\ 
\hline
\multicolumn{10}{c}{Additional SNR candidates} \\
\hline
 811 & 00 42 10.60 & +40 51 49.1 & 6.19   & 3.0e--3$\pm$1.0e--4 & 8.3 &--0.46$\pm$0.46 &  0.48$\pm$0.46 &--0.83$\pm$0.36 &                   \\
 833 & 00 42 14.60 & +40 52 04.7 & 4.30   & 3.2e--3$\pm$5.0e--5 &  26 &  0.73$\pm$0.17&--0.09$\pm$0.17 &--0.84$\pm$0.23 &                   \\
\tablefootmark{c}1121 & 00 43 03.70 & +41 37 17.2 & 4.47   & 5.4e--3$\pm$1.0e--4 &  17 &                &  0.45$\pm$0.24 &  0.14$\pm$0.19 &--0.23$\pm$0.36 \\
1286 & 00 43 42.08 & +41 47 09.5 & 7.42   & 2.2e--3$\pm$5.0e--5 & 8.6 &  0.74$\pm$0.27&--0.38$\pm$0.23 &--0.85$\pm$0.46 &                   \\
\tablefootmark{d}1461 & 00 44 28.62 & +41 49 48.7 & 7.00   & 2.7e--3$\pm$7.0e--5 & 8.7 &  0.63$\pm$0.33&--0.58$\pm$0.29 &                &            \\
\tablefootmark{c}1468 & 00 44 30.56 & +41 23 06.2 & 4.84   & 4.0e--3$\pm$6.0e--5 &  38 &                &  0.82$\pm$0.16 &  0.16$\pm$0.17 &--0.66$\pm$0.30 \\
\tablefootmark{c}1611 & 00 45 18.44 & +41 39 36.3 & 3.42   & 7.5e--3$\pm$8.0e--5 &  94 &                &  0.85$\pm$0.17 &  0.29$\pm$0.11 &--0.13$\pm$0.14 \\
\hline                                    
\end{tabular}
\tablefoot{
Sources in this table have been studied as possible SNR candidates based on optical data 
(see Sect.\,\ref{xdata}).\\
\tablefoottext{a}{Detection likelihood.}
\tablefoottext{b}{Hardness ratios.}
\tablefoottext{c}{Classified as a hard source by SPH11.}
\tablefoottext{d}{Classified as a foreground star candidate by SPH11.}
}
\end{table*}

\begin{figure}
\centering     
\includegraphics[width=0.5\textwidth,clip=]{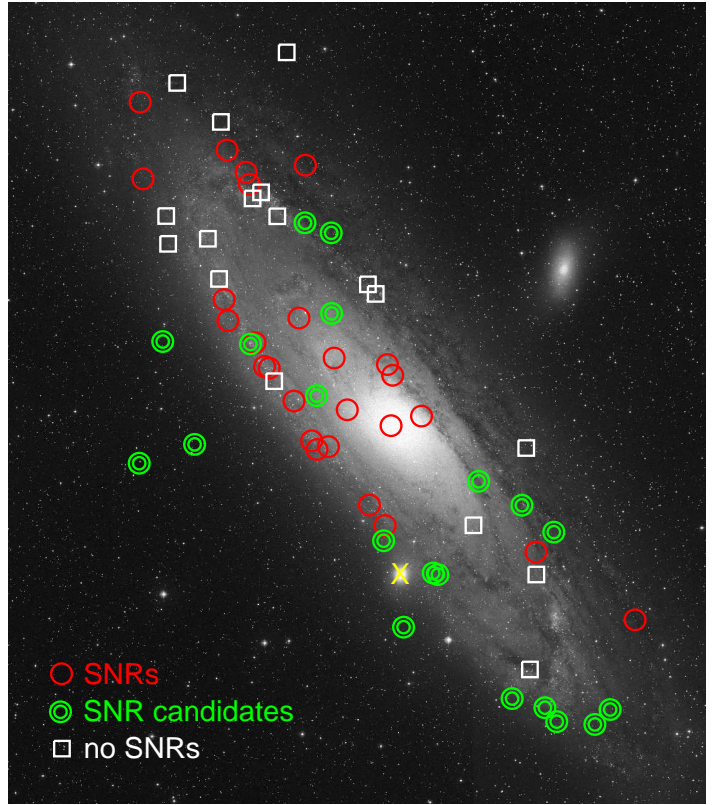}
\caption{
The positions of X-ray SNRs and candidates detected with \xmm\ 
are shown on a Digitised Sky Survey 2 (blue) image. 
The sources are marked with new classifications of this
work (see Sect.\,\ref{classify}).
In addition, the position of the dwarf elliptical M\,32, a satellite 
galaxy of M\,31, is indicated by a yellow cross.
}
\label{dss}
\end{figure}

\begin{figure*}
\centering     
{\bf See 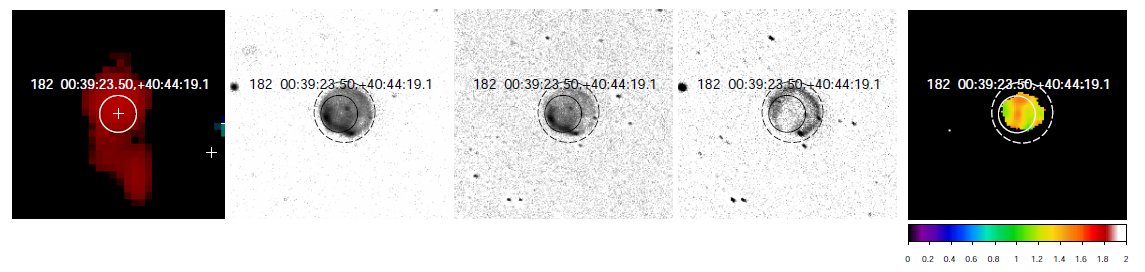.}
\caption{
\xmm\ three colour image (red: 0.2 -- 1.0~keV, green: 1.0 -- 2.0~keV, 
blue: 2.0 -- 4.5~keV), continuum subtracted LGGS H$\alpha$, [\sii], 
[\oiii] images, and [\sii]/\ha\ ratio image with \xmm\ 3$\sigma$ 
positional error circle (solid) for source [SHP11] 182. 
The dashed circle shows the extraction
region for the optical emission. 
The shown area has a size of $\sim$1\arcmin\ $\times$ 1\arcmin. 
The label gives the source number of the \xmm\ catalogue (SPH11) 
followed by the coordinates. The positions of all detected \xmm\ sources 
are marked in the \xmm\ images (left) with crosses.
Appendix \ref{lggs} with images of other sources is
available electronically only.
}
\label{lggs0}
\end{figure*}

\begin{figure*}
\centering     
{\bf See 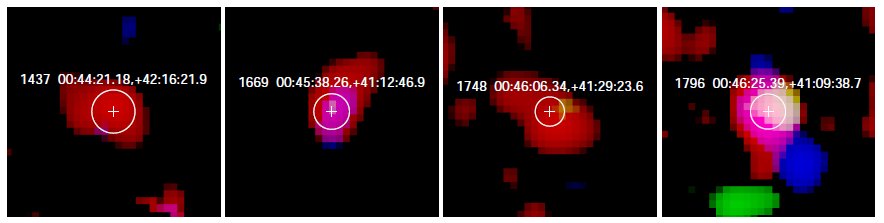.}
\caption{
\xmm\ colour images (red: 0.2 -- 1.0 keV, green: 1.0 -- 2.0 keV, blue: 2.0 -- 4.5 keV)
of sources 1437, 1669, 1748, and 1796, which are located outside the regions covered by 
LGGS (see Sect.\,\ref{optdata}). 
The shown area has a size of $\sim$1\arcmin\ $\times$ 1\arcmin.
}
\label{nolggs}
\end{figure*}

\xmm\ performed a survey of M\,31 as a large programme between June 2006 
and February 2008. The analysis of the entire data 
of this survey including additional archival data taken between June 2000 
and July 2004 is presented by SPH11. 
These observations in total covered the entire $D_{25}$ ellipse 
of M\,31 down to a limiting luminosity of $\sim10^{35}$~erg~s$^{-1}$ in 
the 0.2 -- 4.5~keV band.
A total of 1897 sources were detected in the \xmm\ survey data
(SPH11). 
The sources were identified or classified based on their X-ray hardness
ratios, spatial extent, and variability as well as through cross-correlation
with catalogues in the X-ray, optical, infrared, and radio wavelengths. 
In the end, 25 sources were identified by SPH11 as known SNRs and 31 
additional sources were classified as SNR candidates based on their X-ray 
properties. Their positions are shown on a Digitised Sky Survey (DSS) 2 
blue image in Fig.\,\ref{dss}. 
While the main objective of the work by SPH11 was to create a source 
catalogue of all \xmm\ detections in the field of M\,31, we are 
interested in one particular class of objects, namely SNRs,
which are soft and intrinsically extended sources.
Therefore, we reprocessed and analysed the data using XMMSAS version 
10.0.0 to obtain data products, which are best suited for the study
of SNRs.
We created mosaic images of data taken from all observations 
between 2006 and 2010 in the bands 0.2 -- 1.0~keV, 1.0 -- 2.0~keV, 
and 2.0 -- 4.5~keV.
The images are binned with 2\arcsec\ $\times$ 2\arcsec\ 
pixel size and smoothed with a 
Gaussian with a width of 10\arcsec.
The three colour images created with the three bands zoomed in on each
source are shown in Figs.\,\ref{lggs0}, \ref{nolggs}, and Appendix \ref{lggs}.

The X-ray energy bands used for the detection and calculation
of hardness ratios by SPH11 are 
$B_1$ = 0.2 -- 0.5~keV, $B_2$ = 0.5 -- 1.0~keV, 
$B_3$ = 1.0 -- 2.0~keV, $B_4$ = 2.0 -- 4.5~keV, and $B_5$ = 4.5 -- 12~keV. 
The hardness ratios
and errors are defined as $HR_i = (C_{i+1}-C_i)/(C_{i+1}+C_i)$ and
$EHR_i = 2 \times \sqrt{(C_{i+1} \times EC_i)^2 + (C_{i} \times EC_{i+1})^2)}
/(C_{i+1}+C_i)^2$, $i = 1,2,3,4$. 
$C_i$ and $EC_i$ are the count rates and errors in the energy band $i$.
Table \ref{xlist} lists the SNRs and candidates 
classified by SPH11 along with their positions, count rates, detection 
likelihoods, and hardness ratios. 
If a hardness ratio $HR_1$, $HR_2$, $HR_3$, or $HR_4$ is not given in 
Table \ref{xlist}, it was not calculated
due to poor photon statistics in the corresponding bands. 
The sources in Table \ref{addcandlist} were not classified as SNR candidates
by SPH11. However, the study of optical data while planning
follow-up observations of the \xmm\ LP survey of M\,31 
revealed shell-like optical emission line sources
at the X-ray positions indicating that these sources might be SNRs. 
Therefore, we included these sources also in our work.

\subsection{Optical data}\label{optdata}

\begin{figure}
\centering     
\includegraphics[width=0.38\textwidth,clip=]{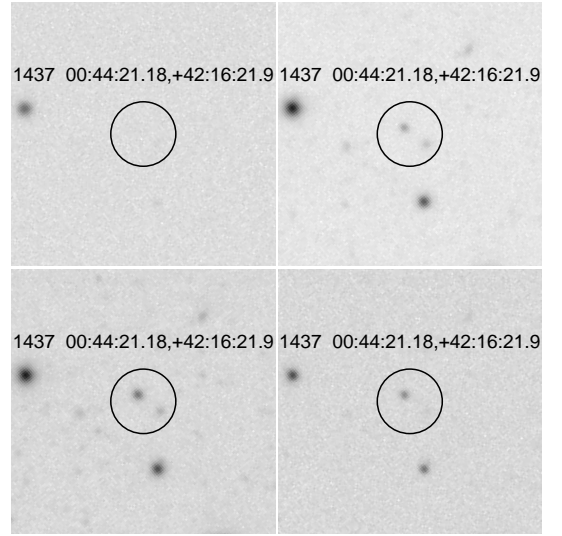}
\caption{
SDSS-III images (upper left: $u$, upper right: $r$, lower left: $i$, lower right: $z$ 
band) of [SPH11] 1437. 
The shown area has a size of $\sim$1\arcmin\ $\times$ 1\arcmin. 
Figure \ref{sdss3} for sources [SPH11] 1669, 1748, and 1796 is
available electronically only.
}
\label{sdss1437} 
\end{figure}

M\,31 was observed during the Local Group Galaxy Survey,
which was carried out with the 4~m telescope at KPNO
\citep{2006AJ....131.2478M}.
Continuum images as well as narrow band \ha, [\sii], and [\oiii] images are 
available for 10 fields, which in total cover the entire galaxy. 
Each field covers $\sim$35\arcmin\ $\times$ 35\arcmin\
and the median seeing of the observations was $\sim$1\arcsec.
Bias-corrected and flat-fielded images can be downloaded
from the LGGS web site\footnote{http://www.lowell.edu/users/massey/lgsurvey.html}.
For the emission line images the photometric uncertainty is 
$\sim$5\% according to the LGGS web page.
We subtracted scaled $R$ band continuum images from the \ha\ and [\sii] images and 
scaled $V$ band continuum images from the [\oiii] images. 

Figures \ref{lggs0} and Appendix \ref{lggs} show 
continuum subtracted LGGS H$\alpha$, [\sii], [\oiii] images, and 
[\sii]/\ha\ ratio image with the \xmm\ 3$\sigma$ positional error circle for 
each source, 
along with the \xmm\ images described in Sect.\,\ref{xdata}.
Four sources ([SPH11] 1437, 1669, 1748, and 1796) are located outside the 
regions 
covered by LGGS. The \xmm\ three colour images of these sources are shown in 
Fig.\,\ref{nolggs}. We used image data from the Data Release 8 of the 
Sloan Digital Sky Survey (SDSS) III\footnote{http://www.sdss3.org/} in order 
to search for possible optical counterparts for these sources. The $u$, $r$, 
$i$, and $z$ band images are shown in Figs.\,\ref{sdss1437} and \ref{sdss3}.
For all sources, we also checked the 
\hst\ archive for optical images in order to search for 
possible counterparts. 
While for most sources the LGGS and SDSS images were sufficient
to identify possible counterparts, in the case of [SPH11] 1505 the
\hst\ image revealed a background galaxy, which is not resolved on
the LGGS images, and thus allowed its classification as a background 
source (see Sect.\,\ref{1505}).

The SNRs and candidates are located in fields F2 to F9 of the LGGS survey.
We first calibrated the images in fields F4, F5, and F6 based on the \ha,
[\sii], and [\oiii] fluxes of [SPH11] 883 (r2-57) and [SPH11] 1040 (r3-84) 
determined by \citet{2004ApJ...615..720W}, allowing us to calculate the flux 
in erg~cm$^{-2}$~s$^{-1}$ from the count rates in the images.
The publication of \citet{2004ApJ...615..720W} also includes the \ha\ and
[\sii] fluxes of [SPH11] 1234. The [\oiii] flux for [SPH11] 1234 was obtained 
from the publication by \citet{1981AJ.....86..989D}.
As the LGGS fields have a small overlap with the adjacent fields (e.g., F3 has
an overlap with F2 and F4), we can, e.g., calibrate F3, if F4 is calibrated 
and there are sources which are observed both in F3 and F4. 
In addition, \ha\ and [\sii] fluxes of sources [SPH11] 
1539 and 1599 could be obtained from \citet{1993A&AS...98..327B}.
By using the fluxes of [SPH11] 1539 and 1599 in F4, which were observed in 
F2, F3, and F4, and comparing the values to their fluxes published by 
\citet{1993A&AS...98..327B}, we calibrated fields 
F2 and F3. For fields F7, F8, F9, we derived the count-rate-to-flux 
conversion 
factors by using other sources that were observed in many fields, starting
with those observed in F6 and F7. 
For fields F8 and F9, however, we were not able to calibrate the [\oiii]
images as there were no sources with significant [\oiii] emission observed
both in F7 and F8. 
This flux calibration yields an uncertainty in the flux in the range
of 3 -- 20\% in most of the fields, but up to 50\% in some fields, in which
there were only a few sources in overlapping areas.

\begin{figure}
\centering     
\includegraphics[width=0.5\textwidth,clip=]{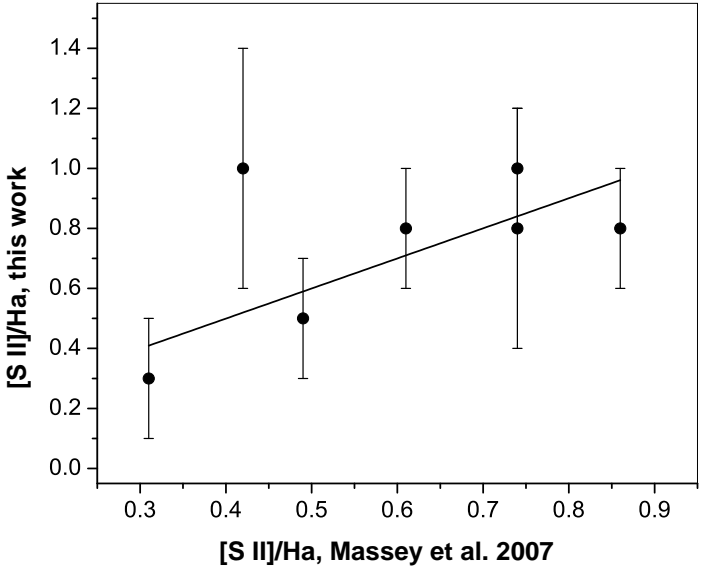}
\caption{
[\sii]/\ha\ flux ratios of sources, for which we computed the flux using 
the LGGS data and there is a corresponding entry in the emission line source
catalogue of \citet{2007AJ....134.2474M}. The fitted line with a slope
of 1.0 is shown.
}
\label{compM07}
\end{figure}

\begin{figure}
\centering     
\includegraphics[height=0.24\textwidth,clip=]{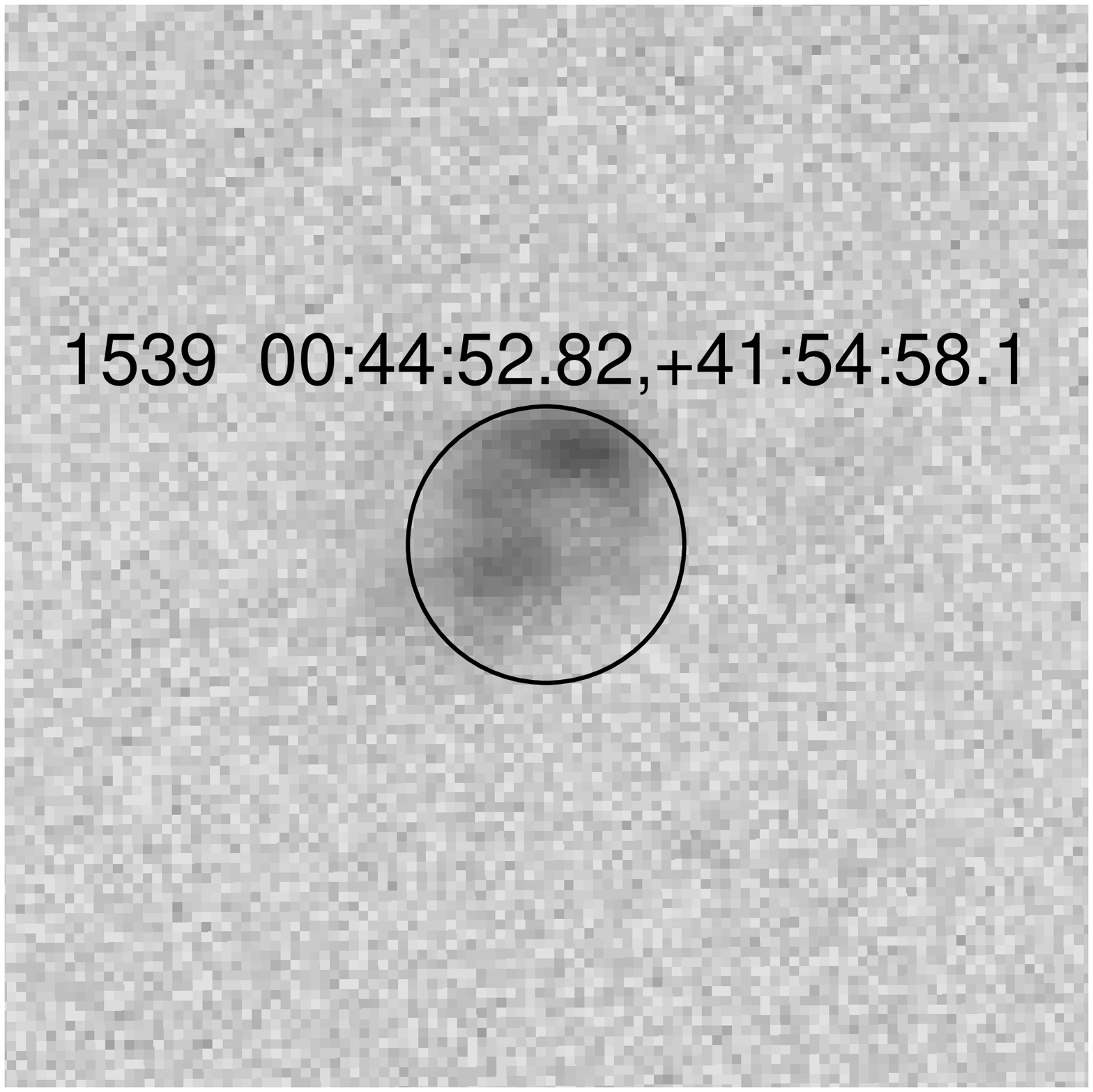}
\includegraphics[height=0.24\textwidth,clip=]{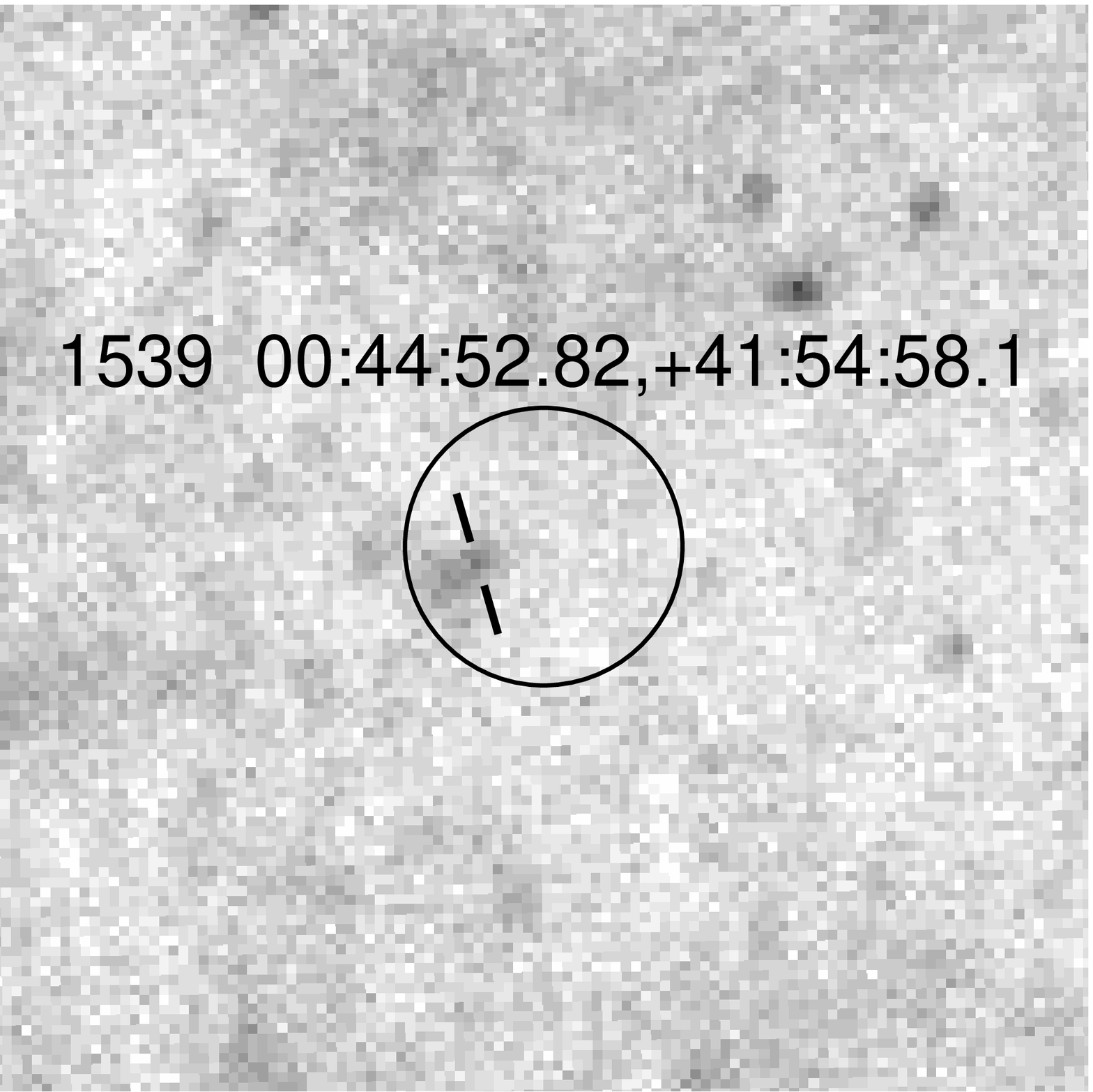}
\caption{
Continuum-subtracted LGGS \ha\ (left) and LGGS $V$ band image (right)
of [SPH11] 1539.
The shown area has a size of $\sim$30\arcsec\ $\times$ 30\arcsec. 
The \xmm\ positional error circle is shown.
The position of LGGS~J004452.98+415457.7 is indicated.
}
\label{v1539}
\end{figure}

We compared our list of sources to the catalogue of emission line objects
by \citet{2007AJ....134.2474M}. We find seven common objects. In Figure 
\ref{compM07}
we plot the [\sii]/\ha\ flux ratios that we derived (see Sect.\,\ref{optflux})
over the values in the \citet{2007AJ....134.2474M} catalogue.
Except for one outlier, the correlation is very good with a coefficient of 
1.002$\pm$0.374 and an intercept of 0.098$\pm$0.235
(Fig.\,\ref{compM07}).
The outlier corresponds to LGGS~J004452.98+415457.7, which 
is the point source visible on the $V$ band 
image inside the positional error circle of the \xmm\ source [SPH11] 1539
(Fig.\,\ref{v1539}, right).
The \ha\ and [\sii] fluxes of the optical source associated to 
[SPH11] 1539 (XMMM31~J004452.82+415458.1) 
was determined in a larger area covering the entire extended optical 
source (Fig.\,\ref{v1539}, left).
This is most likely the reason for the discrepancy between our flux ratio
and that in the catalogue by \citet{2007AJ....134.2474M} corresponding
to the point source.
In conclusion, the agreement between our flux ratios and the
\citet{2007AJ....134.2474M} values for the small sample of common objects
lends credibility to our calibration procedure.

We obtained new optical spectra for three sources, [SPH11] 811, 1156, and 1712, 
in follow-up observations of the \xmm\ LP 
survey undertaken in the period of 2006 -- 2011. The spectra were acquired 
with the 3.5~m telescope at the Apache Point Observatory, located in the
Sacramento Mountains in Sunspot, New Mexico (USA) in order to identify
the optical counterparts of the \xmm\ sources.
These optical spectra will be published in a separate paper (Hatzidimitriou,
Williams et al., in prep.).
In the present work we will only use the results for these three sources to
discuss their nature (see Sect.\,\ref{811}, \ref{1156}, and \ref{1712}).
The telescope was equipped with the Dual Imaging Spectrograph (DIS),
a medium dispersion spectrograph with separate collimators for the
red and blue part of the spectrum and two 2048 $\times$ 1028 E2V CCD cameras,
with the transition wavelength at around 5350 \AA.

\section{Analysis}

\subsection{X-ray spectra}\label{xrayspec}

\begin{figure}
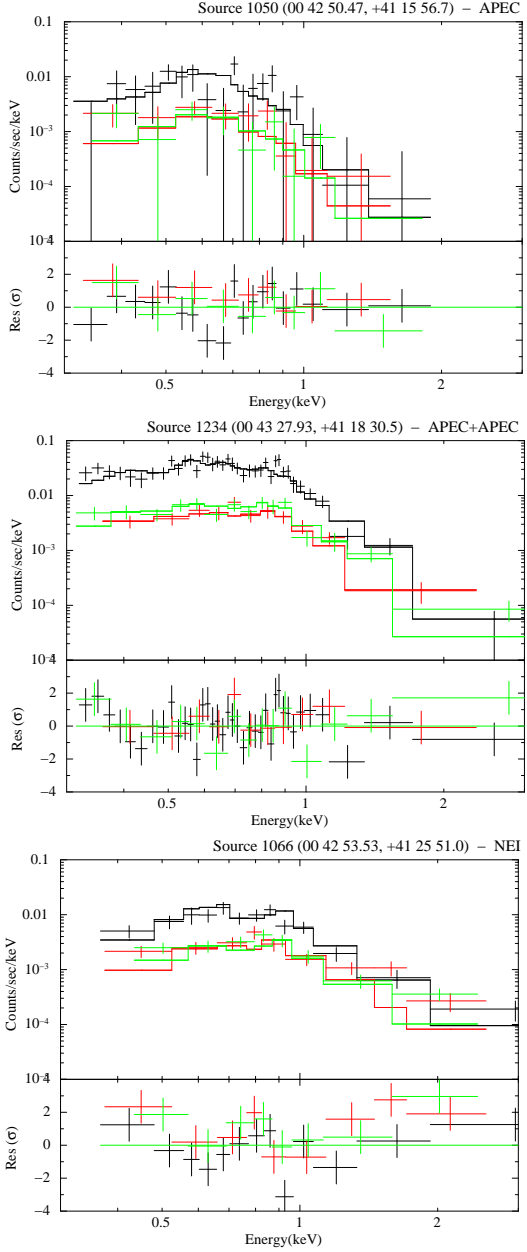

\centering
\includegraphics[width=0.3\textwidth,angle=270,clip=]{04250_411556_apec.ps}
\includegraphics[width=0.3\textwidth,angle=270,clip=]{04327_411830_apec+apec.ps}
\includegraphics[width=0.3\textwidth,angle=270,clip=]{04253_412551_nei.ps}
\caption{
Spectrum of [SPH11] 1050, 1234, and 1066 with an {\tt APEC}, {\tt APEC}+ {\tt APEC},
and NEI model, respectively.
EPIC-pn spectrum with fitted model is shown in black, MOS1 in red, and
MOS2 in green.
}
\label{1050spec}
\end{figure}

\begin{table*}
\caption{
\label{10661234}
Spectral parameters obtained from the fits of the EPIC spectra of [SPH11] 
1050, 1066, and 1234. 
}
\centering
\begin{tabular}{rcccccccc}
\hline\hline
[SPH11] & $N_{\rm H}$ ({\tt TBABS2}) & $kT$ ({\tt APEC}) & $norm$ ({\tt APEC}) & & & Red.\ $\chi^2$ & DOF & $F_{\rm abs}$\tablefootmark{a} \\
ID  & [$10^{21}$~cm$^{-2}$] & [keV] & [cm$^{-5}$] & & & & & [erg~s$^{-1}$~cm$^{-2}$] \\
\hline
1050 & 1 (0 -- 17) & 0.19 (0.04 -- 0.25) & 2e--5 & & & 1.0 & 34 & 1.1e--14 \\ 
1066 & 6 (4 -- 7) & 0.18 (0.14 -- 0.20) & 9e--4 & & & 3.2 & 29 & 3.2e--14 \\ 
1234 & 0 (0 -- 2) & 0.25 (0.05 -- 0.27) & 4e--5 & & & 2.6 & 61 & 5.6e-- 14 \\ 
\hline
[SPH11] & $N_{\rm H}$ ({\tt TBABS2}) & $kT$ ({\tt NEI}) & $\tau$ ({\tt NEI}) & $norm$ ({\tt NEI}) &  & Red.\ $\chi^2$ & DOF & $F_{\rm abs}$\tablefootmark{a}  \\
ID  & [$10^{21}$~cm$^{-2}$] & [keV] & [$10^{10}$~s~cm$^{-3}$] & [cm$^{-5}$] &  & & & [erg~s$^{-1}$~cm$^{-2}$] \\
\hline
1066 & 3 (2 -- 4) & 3.8 (2.8 -- 4.2) & 0.6 (0.4 -- 0.8) & 1e--5 & & 2.3 & 26 & 3.5e--14 \\ 
1234 & 0 (0 -- 1) & 2.2 (1.8 -- 3.8) & 0.9 (0.8 -- 1.2) & 4e--6 & & 1.4 & 60 & 6.2e--14 \\ 
\hline
[SPH11] & $N_{\rm H}$ ({\tt TBABS2}) & $kT$ ({\tt APEC1}) & $norm$ ({\tt APEC1}) &  $kT$ ({\tt APEC2}) & $norm$ ({\tt APEC2}) & Red.\ $\chi^2$  & DOF & $F_{\rm abs}$\tablefootmark{a} \\
ID & [$10^{21}$~cm$^{-2}$] & [keV] & [cm$^{-5}$] & [keV] & [cm$^{-5}$] & & & [erg~s$^{-1}$~cm$^{-2}$] \\
\hline
1234 & 4 (3 -- 6) & 0.23 (0.18 -- 0.25) & 2.8e--4 & 0.04 (0.02 -- 0.05) & 3.9 & 1.1 & 59 & 6.4e--14 \\ 
\hline
\end{tabular}
\tablefoot{
90\% confidence ranges are given in brackets. \\
\tablefoottext{a}{For 0.35 -- 2.0 keV.} 
}
\end{table*}

Out of the 63 sources in Tables \ref{xlist} and \ref{addcandlist}, 
only twelve had enough counts to obtain spectra with good statistics. 
We extracted spectra for these sources and created
corresponding ancillary response files and response matrix files.

We analysed spectra that yielded
more than five bins with more than 20 counts per bin.
The \xmm\ EPIC spectra were all fitted in 
XSPEC with a model that included two absorption components (fixed {\tt TBABS1} 
for the column density in the Milky Way [MW] in the direction of M\,31 of 
$N_{\rm H}$(MW) $= 0.7 \times 10^{21}$ cm$^{-2}$ [Stark et al.\ 
\citeyear{1992ApJS...79...77S}] and a free parameter 
{\tt TBABS2} for additional absorption in M\,31) and a thermal emission 
model. 
To begin with, we used the emission model {\tt APEC}, which assumes
collisional ionisation equilibrium (CIE). For most of the sources, for which
we do not have high statistics to study the effect of non-equilibrium ionisation
(NEI), this model gives us an average temperature of the hot plasma.
The abundances were fixed to solar values \citep{1989GeCoA..53..197A}
in all fits.
The spectra are all fitted relatively well with a low temperature of 
$kT \approx 0.2$~keV. 
The results for the fits using a single {\tt APEC} model component are
listed in the upper part of Table \ref{10661234} for the brightest 
sources [SPH11] 1050, 1066, and 1234. 
For the faintest of the three, 
[SPH11] 1050, the {\tt APEC} model reproduces the spectrum fairly well
with reduced $\chi^2$ = 1.0 (Fig.\,\ref{1050spec}, top).
However, the fits yield no satisfactory results for the brighter sources.
Therefore, we also modeled the spectra with a 
non-equilibrium ionisation model {\tt NEI} or a combination of
two {\tt APEC} models instead of a single {\tt APEC} model
(see Table \ref{10661234}).

The brightest X-ray SNR in M\,31 is [SPH11] 1234 (XMMM31~J004327.93+411830.5).
The spectral analysis revealed that its spectrum consists of at least two
thermal components with temperatures of $kT = 0.04 (0.02 - 0.05)$~keV and 
$kT = 0.23 (0.18 - 0.25)$~keV (Table \ref{10661234} and 
Fig.\,\ref{1050spec}, middle).
It is an SNR first identified in the optical with a nice round
shell clearly visible in \ha, [\sii], and [\oiii]
(see Sect.\,\ref{1234}). As can be seen in the
\ha\ data, the remnant is located in an \hii\ region. The
X-ray spectrum therefore seems to be the superposition of the hot gas
inside an interstellar bubble and the emission of the SNR.

The second brightest source is [SPH11] 969 (XMMM31~J004239.82+404318.8), which
has been suggested to be an SNR candidate based on its soft X-ray spectrum. The 
spectral analysis by SPH11
revealed that its spectrum is best modeled with a non-equilibrium plasma with a 
low temperature $kT$ = 0.2~keV and a strikingly low ionisation timescale 
$\tau = 2 \times 10^{8}$~s~cm$^{-3} \approx 6$~yrs~cm$^{-3}$,
implying a very low density of $n_e = 6 \times 10^{-4}$~cm$^{-3}$
if the SNR had an age of $\sim$10,000~yrs. 
The source is located at the outer rim of the optical disk of M\,31
(green annulus south of M\,32 in Fig.\,\ref{dss}).
Therefore, if this is an SNR it might have been expanding in a very low density 
medium. We tried to fit its spectrum with a model including a power-law component,
which yields an unreasonably high index of $\Gamma$ = 6.0 (5.2 -- 7.0).
Higher statistics X-ray data are necessary to understand the emission of
this source.

The third brightest source [SPH11] 1066 (XMMM31~J004253.53+412551.0) is an
optically identified SNR and has a rather hard spectrum in X-rays
with a high temperature and low ionisation timescale (see Table \ref{10661234}
and Fig.\,\ref{1050spec}, bottom).
This spectrum is consistent with a young SNR.
The statistics are not high enough to verify if there is a non-thermal 
component.


\subsection{Optical flux}\label{optflux}

\begin{table*}
\caption{
\label{snrlist}
X-ray SNRs in M\,31 detected with \xmm. 
}
\centering
\begin{tabular}{rccccc}
\hline\hline
[SPH11] & \ha\ & [\sii] & [\oiii] & [\sii]/\ha\ & $L$ (0.35 -- 2.0~keV)\tablefootmark{a} \\
\multicolumn{1}{c}{ID} & [erg~cm$^{-2}$~s$^{-1}$~arcsec$^{-2}$] & [erg~cm$^{-2}$~s$^{-1}$~arcsec$^{-2}$] & [erg~cm$^{-2}$~s$^{-1}$~arcsec$^{-2}$] & & [erg~s$^{-1}$] \\
\hline
\multicolumn{6}{c}{Known SNRs\tablefootmark{b}} \\
\hline
182   & 1.4e--14$\pm$2.9e--15 & 1.5e--14$\pm$7.5e--15 & N/A                    & 1.1$\pm$0.6 & 2.2e+35 \\
474   & 2.5e--14$\pm$5.2e--15 & 2.6e--14$\pm$1.3e--14 & 2.3e--14$\pm$1.1e--14  & 1.0$\pm$0.5 & 6.9e+35 \\
883   & 9.7e--15$\pm$5.7e--16 & 9.2e--15$\pm$1.9e--15 & 2.4e--16$\pm$2.7e--17  & 0.9$\pm$0.2 & 4.0e+35 \\
1040  & 1.1e--14$\pm$6.5e--16 & 9.0e--15$\pm$1.9e--15 & 1.7e--14$\pm$1.9e--15  & 0.8$\pm$0.2 & 1.0e+36 \\
1050  & 7.5e--15$\pm$4.4e--16 & 7.7e--15$\pm$1.5e--15 &      0.0               & 1.0$\pm$0.2 & 1.2e+36 \\
1066  & 2.2e--14$\pm$1.3e--15 & 1.9e--14$\pm$3.9e--15 & 2.1e--14$\pm$2.3e--15  & 0.9$\pm$0.2 & 3.8e+36 \\
1234  & 5.6e--14$\pm$3.3e--15 & 4.2e--14$\pm$8.7e--15 & 3.7e--14$\pm$4.1e--15  & 0.7$\pm$0.2 & 8.0e+36 \\
1275  & 3.0e--14$\pm$1.7e--15 & 2.8e--14$\pm$5.8e--15 & 5.4e--14$\pm$6.0e--15  & 0.9$\pm$0.2 & 3.1e+36 \\
1291  & 3.1e--17$\pm$1.9e--18 &      0.0              & 1.8e--17$\pm$2.0e--18  & 0.0         & 1.8e+36 \\
1328  & 5.9e--15$\pm$3.4e--16 & 2.5e--15$\pm$5.1e--16 & 1.6e--14$\pm$1.8e--15  & 0.4$\pm$0.1 & 1.2e+36 \\
1351  & 3.8e--14$\pm$2.2e--15 & 3.1e--14$\pm$6.4e--15 & 6.4e--14$\pm$7.2e--15  & 0.8$\pm$0.2 & 6.5e+35 \\
1386  & 5.9e--14$\pm$3.4e--15 & 1.3e--14$\pm$2.7e--15 & 6.3e--15$\pm$7.1e--16  & 0.2$\pm$0.0 & 2.2e+35 \\
1410  & 1.2e--14$\pm$6.8e--16 & 1.1e--14$\pm$2.3e--15 & 7.1e--15$\pm$8.0e--16  & 0.9$\pm$0.2 & 8.5e+35 \\
1497  & 4.7e--14$\pm$2.7e--15 & 2.5e--14$\pm$5.2e--15 & 3.0e--14$\pm$3.4e--15  & 0.5$\pm$0.1 & 5.9e+35 \\
1522  & 2.4e--14$\pm$1.4e--15 & 2.5e--14$\pm$5.1e--15 & 1.2e--14$\pm$1.3e--15  & 1.0$\pm$0.2 & 4.4e+35 \\
1539  & 2.1e--14$\pm$6.4e--15 & 2.1e--14$\pm$4.3e--15 & 2.7e--15$\pm$8.2e--16  & 1.0$\pm$0.4 & 2.1e+35 \\
1587  &      0.0              &      0.0              &      0.0               & 0.0         & 3.6e+35 \\
1593  & 2.6e--14$\pm$7.9e--15 & 2.6e--14$\pm$5.4e--15 & 6.2e--14$\pm$1.9e--14  & 1.0$\pm$0.4 & 1.7e+35 \\
1599  & 5.8e--14$\pm$3.4e--15 & 5.4e--14$\pm$1.1e--14 & 9.7e--14$\pm$2.9e--14  & 0.9$\pm$0.2 & 2.1e+36 \\
1793  & 3.5e--16$\pm$1.1e--16 & 2.5e--16$\pm$5.2e--17 & 2.9e--16$\pm$8.8e--17  & 0.7$\pm$0.3 & 5.1e+35 \\
1805  & 3.2e--17$\pm$9.7e--18 & 9.6e--17$\pm$2.0e--17 & 4.5e--16$\pm$1.3e--16  & 3.0$\pm$1.1 & 8.0e+35 \\
\hline
\multicolumn{6}{c}{Newly confirmed SNRs based on optical and X-ray data} \\ 
\hline
1079  & 3.9e--15$\pm$8.1e--16 & 3.4e--15$\pm$1.7e--15 & 1.4e--14$\pm$7.0e--15  & 0.9$\pm$0.4 & 5.5e+35 \\
1148  & 2.3e--16$\pm$1.3e--17 & 1.8e--16$\pm$3.7e--17 & 1.7e--15$\pm$1.9e--16  & 0.8$\pm$0.2 & 2.9e+35 \\
1370  & 8.2e--17$\pm$4.8e--18 & 6.9e--17$\pm$1.4e--17 & 2.6e--16$\pm$7.9e--17  & 0.8$\pm$0.2 & 3.0e+35 \\
1481  & 7.6e--14$\pm$4.4e--15 & 6.3e--14$\pm$1.3e--14 & 3.7e--14$\pm$4.2e--15  & 0.8$\pm$0.2 & 4.6e+35 \\
1548  & 2.1e--15$\pm$6.4e--16 & 2.3e--15$\pm$4.8e--16 & 1.5e--15$\pm$4.6e--16  & 1.1$\pm$0.4 & 2.1e+35 \\
\hline
\end{tabular}
\tablefoot{
No [\oiii] flux was calculated for sources that were only observed in 
Fields F8 and/or F9 ([SPH11] 182; see Sect.\,\ref{optdata}).\\
\tablefoottext{a}{See Sect.\,\ref{cumxlf}.}
\tablefoottext{b}{SNRs known from literature. The optical fluxes 
and the flux ratios were determined from the LGGS data but were not
used for the classification. }
}
\end{table*}

As mentioned earlier, 
except for four sources, the X-ray SNRs and candidates are located
in regions covered by the LGGS data. For these sources, we determined the
\ha, [\sii], and [\oiii] fluxes and calculated the flux ratio [\sii]/\ha.
The [\sii]/\ha\ flux ratio is a diagnostic tool to distinguish the 
shock-ionised diffuse emission of SNRs from those of photo-ionised \hii\ 
regions or planetary nebulae. 
The radiative shocks in SNRs produce a higher [\sii]/\ha\ ratio, 
typically higher than 0.5 \citep{1993A&AS...98..327B}.
Values around [\sii]/\ha\ = 0.4 -- 0.5 might indicate an SNR
nature, however, shells or filaments in \hii\ regions can 
also reach similar [\sii]/\ha\ ratios. Therefore, we apply the hard limit 
of [\sii]/\ha\ $> 0.5$ for SNRs.
To measure the flux, we looked at the optical images for all sources, 
if available, one by one. If there is optical emission that seems to be
the optical counterpart of the X-ray source the extraction region was
adjusted by eye to cover the entire optical emission.
For sources without obvious optical emission, the size of the 3$\sigma$ error 
circle of the \xmm\ position was used. For instance, the extraction region of
[SPH11] 414 in Fig.\,\ref{lggs}1 was left as it is shown in the figures,
whereas for [SPH11] 182 in Fig.\,\ref{lggs0}, 
the extraction region was increased in size and 
shifted to cover the entire circular optical source seen in all optical 
images (dashed circle).
The background emission was estimated locally for each source.
Based on the optical fluxes and the [\sii]/\ha\ ratio, we confirmed
five sources with significant \ha, [\sii], and [\oiii] emission and an 
[\sii]/\ha\ ratio higher than 0.5 as new SNRs (see Table \ref{snrlist}). 

Another class of extended soft X-ray sources are interstellar bubbles and 
superbubbles in the interstellar medium  of a galaxy, which are formed by 
stellar winds of massive stars. 
Studies of superbubbles in the Large Magellanic Cloud by, e.g.,
\citet{1990ApJ...365..510C} or \citet{2001ApJS..136..119D} have shown 
that the X-ray luminosities of X-ray bright superbubbles is higher than 
what is predicted by theoretical models \citep[e.g., the 
standard model by][]{1977ApJ...218..377W}. This indicates that the energy
sources of the X-ray bright superbubbles are not only the stellar winds
of massive stars, but also supernovae, and thus the shock waves of
SNRs that occurred inside the superbubble.
In contrast to the shock-ionised gas of SNRs, the gas 
in \hii\ regions around such superbubbles is mainly photo-ionised and the 
expansion velocities
are too low to form a radiative shock. Therefore, the [\sii]/\ha\ 
ratio is lower in superbubbles than in SNRs. Some sources in our list have 
optical counterparts with a large extent and low [\sii]/\ha\ ratio ($\sim$0.3).
These sources are most likely superbubbles or SNRs in a superbubble. 
We want to point out that sources classified as SNR candidates in the catalogue 
of SPH11 can as well be superbubbles since the classification was
based on X-ray hardness ratios, which will indicate a soft source for both 
superbubbles and SNRs.

\section{New classifications}\label{classify}

All sources in Tables \ref{xlist} and \ref{addcandlist}
have been studied in detail and are (re-)classified as SNR, SNR candidates, or rejected
(see Tables \ref{snrlist}, \ref{candlist}, and \ref{nosnrlist}).

Here, we summarise the X-ray and optical results for each source
and its identification. 
In the SPH11 catalogue, following criteria were applied to identify
SNRs and candidates:
\begin{enumerate}
\item Soft sources with $HR_1 > -0.1$, $HR_2 + EHR_2 <-0.2$ are
classified as SNR candidates if no foreground star was detected at the 
X-ray position.
\item A source that fulfilled criterion 1 and had a known SNR 
as a counterpart was identified as an SNR.
\item A source that did not fulfill criterion 1 but had an SNR 
counterpart in radio or optical and showed no significant variability 
was classified as an SNR candidate.
\end{enumerate}


We have analysed the LGGS \ha, [\sii], and [\oiii] data to obtain
the optical emission line flux of the sources and to study their morphology.
%
In the following, we will discuss for each source 1) why the source
was classified as an SNR or a candidate by SPH11 and  
if there are any interesting details
in the X-ray image or spectra, 2) its optical properties, 
and 3) a revised classification based on these new results.
Specifically, we have 
\begin{itemize}
\item {\it confirmed an \xmm\ source as an SNR} if the X-ray 
source is soft (i.e., fulfills criterion
1 of SPH11), there is no foreground star at its position, and
is has either a radio counterpart classified as an SNR
or an optical counterpart with [\sii]/\ha\ $> 0.5$,
\item {\it newly classified an \xmm\ source as an SNR} if
its spectrum is soft, there is no foreground star at its position, 
and the optical counterpart incidates an SNR ([\sii]/\ha\ $> 0.5$),
\item {\it classified an \xmm\ source as an SNR candidate}
if a source has a soft X-ray spectrum, no foreground star
as counterpart, and neither optical nor radio counterpart 
indicative of an SNR, \\
OR a source has a soft X-ray spectrum, a foreground star at its position,
but also an optical counterpart with [\sii]/\ha\ $> 0.5$,\\
OR a source had been identified as an SNR before and has a soft
X-ray spectrum, but a possible foreground star together with
diffuse optical emission was found in this work, \\
OR a source has a hard X-ray spectrum and a radio counterpart, \\
OR a hard source has an optical counterpart with [\sii]/\ha\ $> 0.5$.
\end{itemize}

The new classifications are presented in the following. The details
on the X-ray and optical properties for the sources, for which the 
classication remains unchanged, are available electronically in 
App.\,\ref{unchanged}.
We assume a distance of 744~kpc to M\,31 \citep{2010A&A...509A..70V}.
At this distance 1\arcsec\ corresponds to 3.6~pc.
We use the following abbreviations for publications, which we refer to
in this section:
\citet[BA64]{1964ApJ...139.1027B},
\citet[B90]{1990ApJS...72..761B},
\citet[BW93]{1993A&AS...98..327B},
\citet[DDB80]{1980A&AS...40...67D},
\citet[DKG04]{2004ApJ...610..247D},
\citet[GLG04]{2004ApJS..155...89G},
\citet[GLG05]{2005ApJS..159..242G},
\citet[KGP02]{2002ApJ...577..738K},
\citet[MPV95]{1995A&AS..114..215M},
\citet[O06]{2006ApJ...643..844O},
\citet[PAV78]{1978A&AS...31..439P},
\citet[PFH05]{2005A&A...434..483P},
\citet[SBK09]{2009A&A...495..733S}, 
\citet[WB92]{1992A&AS...92..625W},
\citet[WGK04]{2004ApJ...609..735W}, and
\citet[WSK04]{2004ApJ...615..720W}.

\subsection{New SNRs}

\subsubsection{[SPH11] 1079: XMMM31~J004255.50+405946.4}

[SPH11] 1079 is a soft X-ray source with no foreground star at its
position. 
Therefore, it was classified as an SNR candidate by SPH11.

There is a diffuse optical source located at the X-ray position, visible in 
all three \ha, [\sii], and [\oiii] images (Fig.\,\ref{lggs}5). The flux ratio 
is [\sii]/\ha\ = $0.9\pm0.4$, indicating that it is an SNR. The extent of
the diffuse emission is $\sim$9\arcsec. It has a bright concave 
arc-like structure in the south-west.
Based on the X-ray hardness ratios and the optical flux ratio,
we classify this source as a new SNR.

\subsubsection{[SPH11] 1148: XMMM31~J004308.85+410305.4}

[SPH11] 1148 is a soft X-ray source. 
As no foreground star was found at its position, 
it was classified as an SNR candidate by SPH11.

The flux ratio derived from the LGGS data is [\sii]/\ha\ = $0.8\pm0.2$ 
(Table \ref{snrlist}) suggesting that it is an SNR.
It is very faint and diffuse on the optical emission line images with an
extent of $\sim$12\arcsec\ and 
patchy structures (Fig.\,\ref{lggs}5). This
morphology is similar to that of [SPH11] 1282 and 1328.
Based on the X-ray colour and the optical line flux ratio we classify 
this source as an SNR.

\subsubsection{[SPH11] 1370: XMMM31~J004404.55+415806.5}

[SPH11] 1370 is a soft X-ray source
without a foreground star at its position, which therefore
was classified as an SNR candidate by SPH11.

We detected very faint \ha, [\sii], and [\oiii] emission in the LGGS data 
with a flux ratio of [\sii]/\ha\ = $0.8\pm0.2$ (see Table \ref{snrlist}),
which is hardly visible on the images (Fig.\,\ref{lggs}7).
This optical line flux ratio indicates that this source is indeed an SNR.
Therefore, the source is now classified as an SNR.

\subsubsection{[SPH11] 1481: XMMM31~J004434.90+412512.7}

[SPH11] 1481 is a faint, soft source 
with no foreground star at its position. Based on its hardness ratios
it was classified as an SNR candidate by SPH11. 

The X-ray source is coincident with the optical source [BW93] K490A located 
slightly to the north, which had been suggested to be an SNR candidate 
(Fig.\,\ref{lggs}8).
The [\sii]/\ha\ flux ratio derived from the LGGS
data is 0.8$\pm$0.2 and indicates that the source is an SNR.
The optical source is diffuse and has an extent of $\sim$20\arcsec. 
It seems to be located in a complex of \hii\ regions.
The \xmm\ source is also located in a region with extended soft X-ray 
emission and was not detected right at the position of the optical SNR.
If the X-ray detection had been placed at the position of the
optical SNR, the X-ray count rates would have most likely been 
higher (see Fig.\,\ref{lggs}8). Therefore, the count rates in the
catalogue of SPH11 are rather underestimated.
Based on the X-ray and optical properties, we classify this source as a 
new SNR.

\subsubsection{[SPH11] 1548: XMMM31~J004455.73+415655.2}

[SPH11] 1548 is a faint, soft X-ray source
which has been suggested as an SNR candidate by SPH11
based on its hardness ratios and the lack of a foreground star as
likely counterpart.

There is a faint arc-like structure in the optical with an
extent of $\sim$8\arcsec, which is visible in \ha, [\sii], and [\oiii] 
at the X-ray position (Fig.\,\ref{lggs}10). 
The [\sii]/\ha\ flux ratio of 1.1$\pm$0.4
suggests that this object is an SNR.
Therefore, we identify this source as a new SNR.

\subsection{New SNR candidates}

\begin{table*}
\caption{
\label{candlist}
X-ray SNR candidates in M\,31 detected with \xmm. 
}
\centering
\begin{tabular}{rccccc}
\hline\hline
[SPH11] & \ha\ & [\sii] & [\oiii] & [\sii]/\ha\ & $L$ (0.35 -- 2.0~keV)\tablefootmark{a} \\
\multicolumn{1}{c}{ID} & [erg~cm$^{-2}$~s$^{-1}$~arcsec$^{-2}$] & [erg~cm$^{-2}$~s$^{-1}$~arcsec$^{-2}$] & [erg~cm$^{-2}$~s$^{-1}$~arcsec$^{-2}$] & & [erg~s$^{-1}$] \\
\hline
263   & 1.4e--14$\pm$2.9e--15 & 1.1e--14$\pm$5.5e--15 & N/A                    & 0.8$\pm$0.4 &  4.0e+35\\
294   & 1.4e--16$\pm$2.9e--17 & 4.2e--17$\pm$2.1e--17 & N/A                    & 0.3$\pm$0.2 &  2.1e+36\\
414   &      0.0              & 2.3e--17$\pm$1.2e--17 & N/A                    & 0.0         &  4.4e+35\\
419   &      0.0              & 5.6e--18$\pm$2.8e--18 & 2.5e--16$\pm$1.3e--16  & 0.0         &  2.3e+35\\
441   & 2.1e--16$\pm$4.3e--17 & 7.5e--17$\pm$3.7e--17 & N/A                    & 0.4$\pm$0.2 &  4.0e+35\\
521   & 1.1e--12$\pm$2.3e--13 & 2.8e--13$\pm$1.4e--13 & 8.0e--13$\pm$4.0e--13  & 0.3$\pm$0.2 &  2.4e+35\\
560   & 1.8e--16$\pm$3.7e--17 & 5.2e--17$\pm$2.6e--17 & N/A                    & 0.3$\pm$0.2 &  6.1e+35\\
668   & 1.5e--14$\pm$3.1e--15 & 1.3e--14$\pm$6.5e--15 & 3.1e--14$\pm$1.6e--14  & 0.9$\pm$0.5 &  1.1e+36\\
811   & 5.4e--15$\pm$1.1e--15 & 7.8e--15$\pm$3.9e--15 & 2.6e--14$\pm$1.3e--14  & 1.4$\pm$0.8 &  3.3e+35\\
833   & 3.3e--15$\pm$6.8e--16 & 4.7e--15$\pm$2.3e--15 & 6.4e--15$\pm$3.2e--15  & 1.4$\pm$0.7 &  3.5e+35\\
969   &      0.0              & 5.6e--18$\pm$2.8e--18 & 3.4e--17$\pm$1.7e--17  & 0.0         &  7.9e+36\\
1083  &      0.0              &      0.0              & 8.1e--17$\pm$4.0e--17  & 0.0         &  2.2e+35\\
1282  & 4.1e--15$\pm$2.4e--16 & 1.9e--15$\pm$3.9e--16 & 3.9e--15$\pm$1.2e--15  & 0.5$\pm$0.1 &  2.6e+35\\
1286  & 4.0e--16$\pm$2.3e--17 & 2.1e--16$\pm$4.3e--17 & 1.3e--15$\pm$4.0e--16  & 0.5$\pm$0.1 &  2.4e+35\\
1332  & 2.2e--16$\pm$1.3e--17 &      0.0              & 2.3e--16$\pm$2.5e--17  & 0.0         &  5.1e+35\\
1372  & 6.2e--14$\pm$3.6e--15 & 1.6e--14$\pm$3.3e--15 & 1.1e--14$\pm$3.3e--15  & 0.3$\pm$0.1 &  2.2e+35\\
1535  & 2.8e--14$\pm$1.6e--15 & 2.9e--14$\pm$6.0e--15 & 6.2e--14$\pm$6.9e--15  & 1.0$\pm$0.2 &  1.6e+36\\
1669  &                       &                       &                        &             &  3.7e+35\\
1748  &                       &                       &                        &             &  2.6e+35\\
1796  &                       &                       &                        &             &  1.1e+36\\
\hline
\end{tabular}
\tablefoot{
No [\oiii] flux was calculated for sources that were only observed in 
Fields F8 and/or F9 ([SPH11] 263, 294, 414, 441, and 560; see 
Sect.\,\ref{optdata}). [SPH11] 1437, 1669, 1748, and 1796 are located 
outside the LGGS fields.\\
\tablefoottext{a}{See Sect.\,\ref{cumxlf}.}
}
\end{table*}

\begin{figure}
\centering     
{\bf See 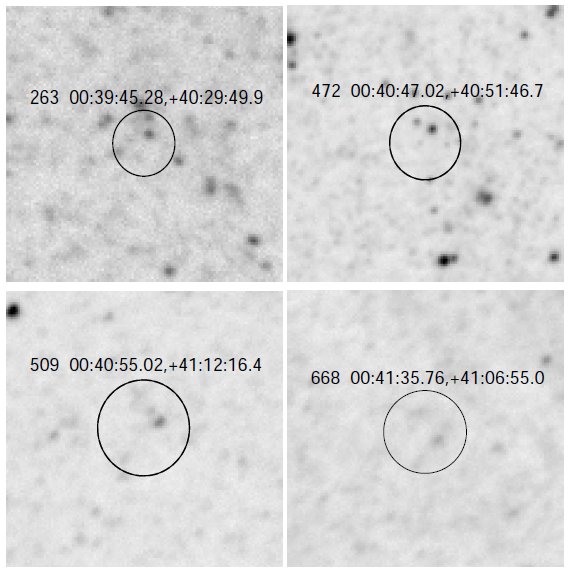 and 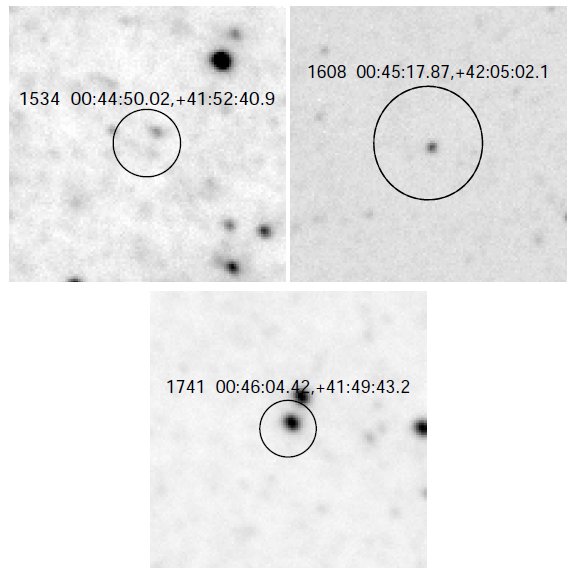.}
\caption{
LGGS $R$ band images of [SPH11] 263, 472, 509, 1534, 1608, and 1741.
The shown area has a size of $\sim$30\arcsec\ $\times$ 30\arcsec. 
The \xmm\ positional error circle is shown.
}
\label{rband}
\end{figure}

\subsubsection{[SPH11]  263: XMMM31~J003945.28+402949.9}

[SPH11] 263 
is a soft X-ray source with hardness ratios 
indicative of an SNR. 
The X-ray source is coincident with the optical SNR [BA64] 339 or 
[PAV78] 80 and was therefore identified as an SNR by PFH05 and
SPH11. 

The optical line emission of the LGGS images (Fig.\,\ref{lggs}1) shows a 
ring-like structure with an extent of $\sim$15\arcsec, 
which has a flux ratio of [\sii]/\ha\ = $0.8\pm0.4$ at 
the position of the X-ray source, and an additional diffuse emission
to the west with a lower flux ratio ([\sii]/\ha\ $\approx 0.4$).
Therefore, it might be an SNR embedded in an interstellar bubble.

The LGGS $B$, $V$, $R$, and $I$ band images
show several faint point-like 
sources at the position of the X-ray source, two of which appear very red 
and may be late-type foreground stars. These stars might contribute to the 
detected X-ray emission (Fig.\,\ref{rband}, top left). 
Due to this ambiguity we change the classification of the X-ray source 
to an SNR candidate. Only X-ray images with higher spatial resolution 
than \xmm\ will allow to resolve the contributions from the different 
sources of X-ray emission.

\subsubsection{[SPH11]  668: XMMM31~J004135.76+410655.0}

[SPH11] 668 
is coincident with the optical SNR [DDB80] 1-11 ([BA64] 416),  
which is also a radio source ([B90] 14). 
The X-ray source has also been detected with \chandra\ ([DKG04] s1-42).
Although the hardness ratio $HR_2 = -0.30\pm0.15$ does not fulfill the
SNR criterion, PFH05 and SPH11 classified the X-ray source as an SNR
based on its positional coincidence with the optical SNR. 

From the narrow-band LGGS images we determine an [\sii]/\ha\ flux ratio 
of $0.9\pm0.5$. The optical SNR has a diameter of $\sim$8\arcsec\ and
is brighter in the north and southwest (Fig.\,\ref{lggs}3).

There is a faint star to the southwest
inside the error radius of the \xmm\ position
(Fig.\,\ref{rband}, upper middle right).
We cannot rule out that the X-ray source is a foreground star. 
X-ray data with higher spatial resolution are necessary to separate 
the emission components. Therefore, we classify the \xmm\ source as
an SNR candidate.

\subsubsection{[SPH11]  811: XMMM31~J004210.60+405149.1}\label{811}

[SPH11] 811 was not classified as an SNR candidate by SPH11.
It is coincident with source [PFH05] 224, which was classified
as a candidate for a super-soft source.
However, a look at DSS images in the course of the preparation for 
optical follow-up observations revealed a ring-like structure.

A circular shell with a diameter of $\sim$12\arcsec,
which is brighter in the east, is clearly visible in the 
\ha, [\sii], and [\oiii] images of the LGGS (Fig.\,\ref{lggs}3). 
The flux ratio of the shell is [\sii]/\ha\ = 1.4$\pm$0.8 supporting its
SNR nature.

\begin{figure}
\centering     
\includegraphics[height=0.24\textwidth,clip=]{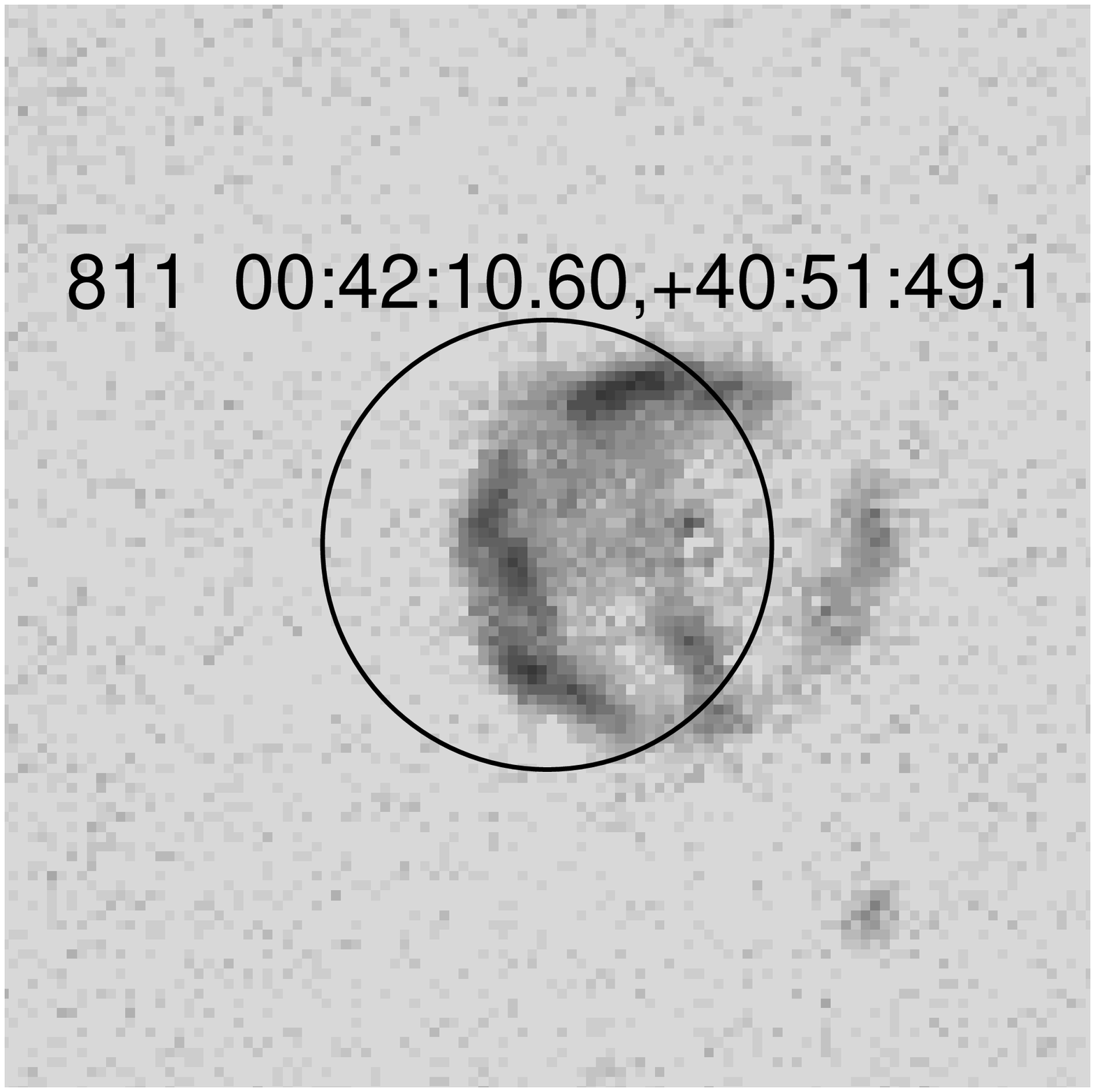}
\includegraphics[height=0.24\textwidth,clip=]{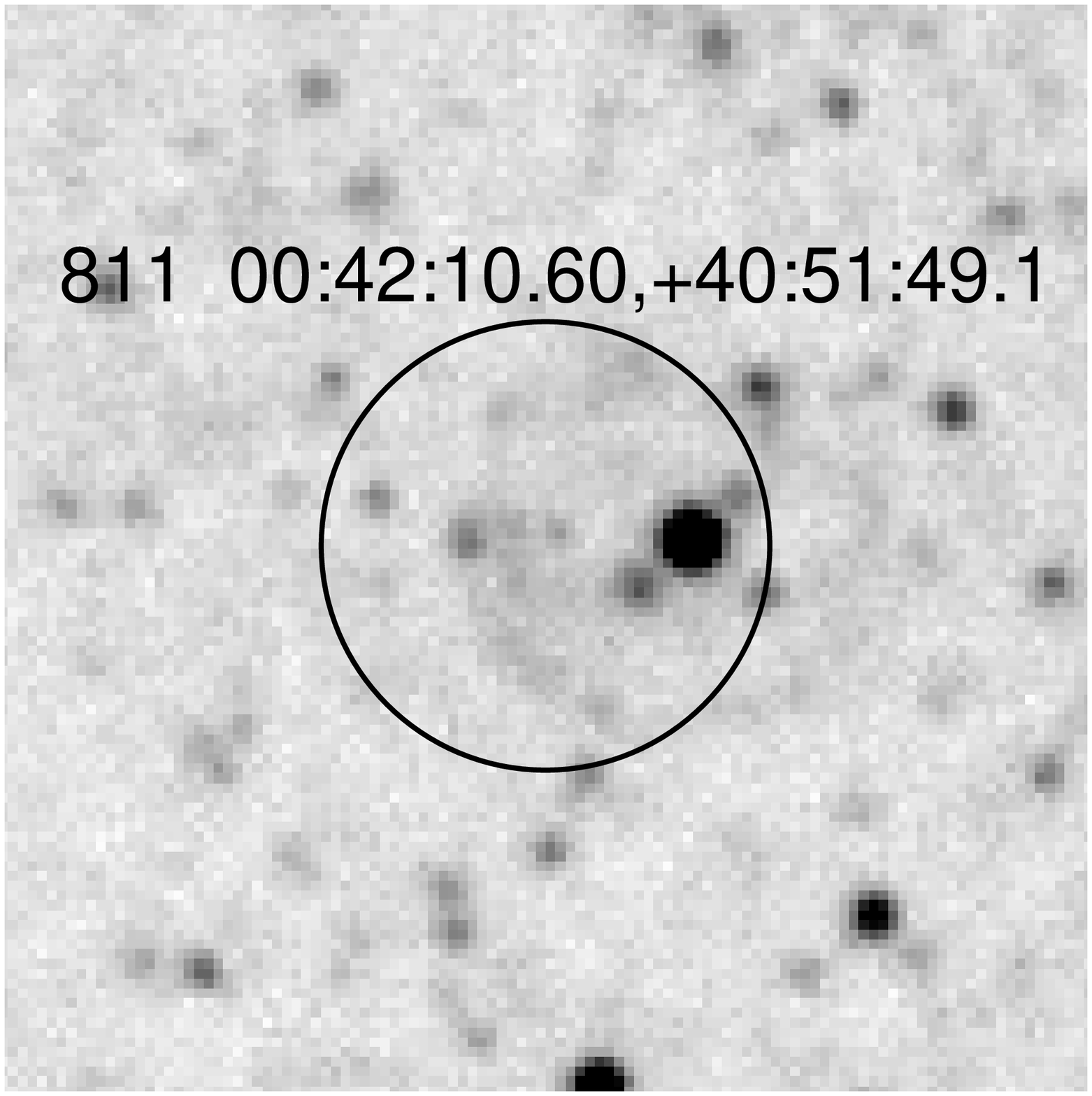}
\caption{
Continuum-subtracted LGGS \ha\ (left) and LGGS $U$ band image (right) 
of [SPH11] 811.
The shown area has a size of $\sim$30\arcsec\ $\times$ 30\arcsec. 
The \xmm\ positional error circle is shown.
}
\label{u811}
\end{figure}

In addition, there is a spoke-like 
feature extending from the centre to the south (Fig.\,\ref{u811}, left). 
On the LGGS $U$, $B$, and $V$ images there is a bright point-like 
source at the centre of the ring with two additional fainter point-like
structures northwest and southeast of the bright source (Fig.\,\ref{u811}, 
right). These three sources form a line almost perpendicular to the 
spoke seen in the emission line images.
We have derived an [\sii]/\ha\ flux ratio of $1.15\pm0.11$ from new optical 
follow-up observations (Hatzidimitriou et al., in prep.), which is in good 
agreement with the flux ratio from the LGGS data. The spectrum also shows 
weak Balmer lines and \hei\ lines in absorption, indicative of an early-type 
star. Therefore, the optical source is most likely a composite of an SNR and 
an OB star in M\,31.

As the X-ray hardness ratios are not indicative of an SNR
($HR_1 = -0.46\pm0.46$, $HR_2 = 0.48\pm0.46$, $HR_3 = -0.83\pm0.36$),
but the optical source is most likely an SNR, 
we classify the \xmm\ source [SPH11] 811 as an SNR candidate.

\subsubsection{[SPH11]  833: XMMM31~J004214.60+405204.7}

[SPH11]  833 has hardness ratios not typical of SNRs.
It is coincident with [PFH05] 234. We have identified
diffuse \ha, [\sii], and [\oiii] emission and thus regarded the source as an SNR
candidate. 

An extended emission line source elongated in the northeast to southwest 
direction ($\sim$9\arcsec\ $\times$ 2\arcsec)
is found on the \ha, [\sii], and [\oiii] LGGS images (Fig.\,\ref{lggs}3). 
The flux ratio of 
[\sii]/\ha\ = 1.4$\pm$0.7 suggests that the source is an SNR.
Therefore, we classify this source as an SNR candidate.

\subsubsection{[SPH11] 1286: XMMM31~J004342.08+414709.5}

[SPH11] 1286 is a soft source, 
which marginally did not fulfill the criterion of
$HR_2 + EHR_2 < -0.2$. Diffuse \ha\ emission was found in the optical.

The LGGS images show a faint diffuse source in \ha, [\sii], and [\oiii] 
(Fig.\,\ref{lggs}6).
The [\sii]/\ha\ flux ratio is $0.5\pm0.1$. This value is at the border between
SNRs and \hii\ regions, thus suggesting an SNR but not clearly ruling out
the \hii\ region identification.  
Therefore, we classify this source as an SNR candidate.

\subsubsection{[SPH11] 1372: XMMM31~J004404.71+414846.7}\label{1372}

[SPH11] 1372 is a faint, soft X-ray source
coincident with an extended optical source [WB92] 280, which was suggested 
as a possible SNR candidate by [MPV95] (source 3-072).
It was therefore classified as an SNR by PFH05 and SPH11.

The optical source is also visible in the LGGS images (Fig.\,\ref{lggs}6). 
The flux ratio derived from the LGGS data is [\sii]/\ha\ = 0.3$\pm$0.1, 
more typical for an \hii\ region than for an SNR. The source is highly extended 
($\sim$30\arcsec) and consists of an extended round region in the north and an 
additional arc-like structure in the southeast.
Therefore, we classify the X-ray source as an SNR candidate, probably located
in an \hii\ region.

\subsubsection{[SPH11] 1535: XMMM31~J004451.07+412905.9}

\begin{figure}
\centering     
\includegraphics[width=0.24\textwidth,clip=]{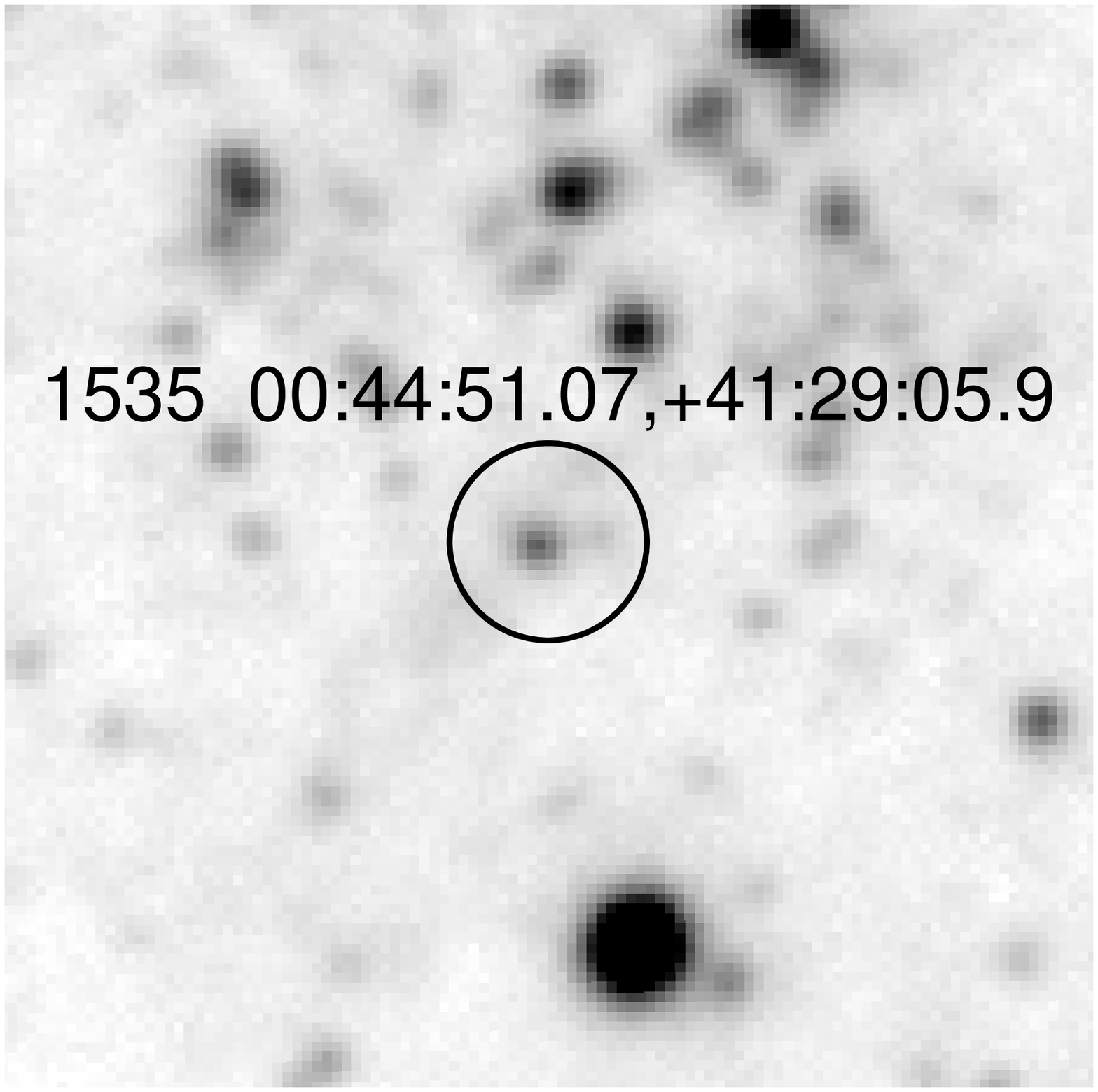}
\includegraphics[width=0.24\textwidth,clip=]{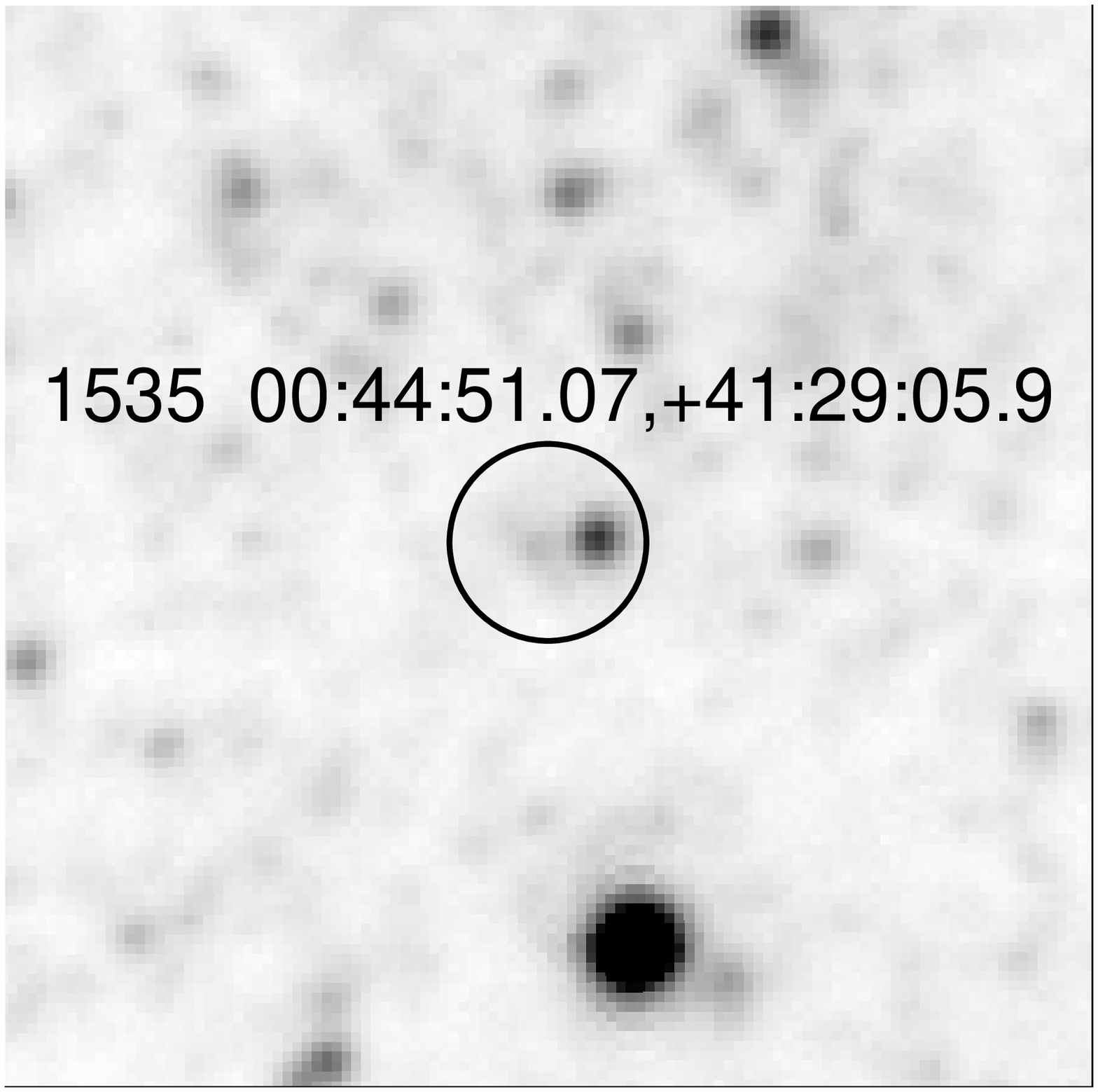}
\caption{
LGGS $V$ (left) and $I$ (right) band images of [SPH11] 1535.
The shown area has a size of $\sim$30\arcsec\ $\times$ 30\arcsec. 
The \xmm\ positional error circle is shown.
}
\label{vi1535}
\end{figure}

[SPH11] 1535
is coincident with the optical SNR [BW93] K583 and its radio 
counterpart and was thus identified as an SNR by PFH05 and SPH11.

The optical SNR is relatively compact with an extent of $\sim$8\arcsec\ and 
an arc-like structure, which is open to the southwest (Fig.\,\ref{lggs}9). 
It has a flux ratio typical for SNRs 
([\sii]/\ha\ = 1.0$\pm$0.2). The SNR seems to be embedded in a shell-like
\hii\ region extending to the northeast, best seen in the \ha\ image.

A closer look at the LGGS $V$, $R$, $I$ images has revealed
two star-like objects within the X-ray error circle, i.e., in the projected
interior of the optical SNR (Fig.\,\ref{vi1535}).  
The object in the centre of the X-ray error circle is a star in M\,31 
with a $V$ magnitude of 19.43
listed as source D31~J004451.1+412905.6 in the catalogue of the DIRECT Project 
\citep{2001AJ....122.1383M}. 
The redder object to the west is 2MASS~00445095+4129058 with a $J$ magnitude of 
16.063, listed in the 2MASS All-Sky Catalog of Point Sources 
\citep{2003yCat.2246....0C}.

The X-ray emission of the SNR is most likely contaminated by
the emission of these stellar objects. X-ray data with higher spatial resolution
are necessary to separate the emission components.
Therefore, we classfiy this \xmm\ source as an SNR candidate.

\subsection{Sources which are classified as no SNR candidates}

\begin{table*}
\caption{
\label{nosnrlist}
\xmm\ sources from SPH11, which are no longer classified
as X-ray SNRs or SNR candidates after new analysis.
}
\centering
\begin{tabular}{rccccll}
\hline\hline
[SPH11] & \ha\ & [\sii] & [\oiii] & [\sii]/\ha\ & Class.\ by & New \\
\multicolumn{1}{c}{ID}  & [erg~cm$^{-2}$~s$^{-1}$~arcsec$^{-2}$] & [erg~cm$^{-2}$~s$^{-1}$~arcsec$^{-2}$] & [erg~cm$^{-2}$~s$^{-1}$~arcsec$^{-2}$] & & SPH11\tablefootmark{a} & Class.\tablefootmark{b} \\
\hline
472   & 2.2e--16$\pm$4.5e--17 & 1.4e--16$\pm$7.0e--17 &      0.0               & 0.6$\pm$0.3 & $<$SNR$>$ & $<$fg star$>$\\
500   & 1.7e--14$\pm$3.5e--15 & 5.4e--15$\pm$2.7e--15 & 3.8e--15$\pm$1.9e--15  & 0.3$\pm$0.2 & $<$SNR$>$ & $<$hard$>$\\
509   & 2.3e--17$\pm$4.8e--18 & 6.4e--17$\pm$3.2e--17 & 0.0e-+00$\pm$1.6e--17  & 2.8$\pm$1.5 & $<$SNR$>$ & $<$fg star$>$\\
682   & 1.2e--17$\pm$2.5e--18 & 3.6e--17$\pm$1.8e--17 &      0.0               & 0.0         & $<$SNR$>$ & \\ 
1121  & 4.6e--14$\pm$2.7e--15 & 9.6e--15$\pm$2.0e--15 & 5.0e--15$\pm$5.6e--16  & 0.2$\pm$0.1 & $<$hard$>$ & $<$hard$>$\\
1156  & 2.2e--15$\pm$1.3e--16 & 1.8e--15$\pm$3.7e--16 & 1.0e--15$\pm$1.1e--16  & 0.8$\pm$0.2 & $<$SNR$>$ & $<$hard$>$\\
1437  &                       &                       &                        &             & $<$SNR$>$ &           \\ 
1461  & 5.0e--14$\pm$2.9e--15 & 9.3e--15$\pm$2.0e--15 & 4.3e--15$\pm$1.3e--15  & 0.2$\pm$0.1 & $<$fg star$>$ & $<$fg star$>$\\
1468  & 5.1e--15$\pm$2.9e--16 & 2.1e--15$\pm$4.3e--16 & 4.4e--16$\pm$4.9e--17  & 0.4$\pm$0.1 & $<$hard$>$ & $<$hard$>$\\
1505  & 2.8e--16$\pm$8.5e--17 & 0.0                   & 2.8e--16$\pm$8.5e--17  & 0.0         & $<$SNR$>$ & $<$AGN$>$\\
1534  & 8.4e--16$\pm$2.5e--16 &      0.0$\pm$5.9e--17 & 7.2e--16$\pm$2.2e--16  & 0.0         & $<$SNR$>$ & $<$fg star$>$\\\
1608  & 2.3e--16$\pm$7.0e--17 &      0.0              &      0.0$\pm$7.9e--18  & 0.0         & $<$SNR$>$ & $<$fg star$>$ \\
1611  & 1.8e--14$\pm$1.0e--15 & 8.2e--15$\pm$1.7e--15 & 2.6e--14$\pm$7.9e--15  & 0.5$\pm$0.1 & $<$hard$>$ & $<$hard$>$\\
1637  & 1.6e--17$\pm$4.9e--18 & 0.0                   & 0.0                    & 0.0         & $<$SNR$>$ & $<$hard$>$\\
1712  & 1.8e--13$\pm$5.5e--14 & 9.4e--14$\pm$2.0e--14 & 3.4e--14$\pm$1.0e--14  & 0.5$\pm$0.2 & $<$SNR$>$ & \\
1732  & 7.2e--15$\pm$4.2e--16 & 2.4e--15$\pm$4.9e--16 & 2.1e--15$\pm$6.4e--16  & 0.3$\pm$0.1 & $<$SNR$>$ & $<$hard$>$\\
1741  & 1.6e--15$\pm$4.9e--16 &      0.0              &      0.0$\pm$3.2e--17  & 0.0         & $<$SNR$>$ & $<$fg star$>$\\
\hline
\end{tabular}
\tablefoot{
\tablefoottext{a}{Classification by SPH11. fg star: foreground star.
The brackets $< >$ indicate that these are candidates and the classifications 
hence need to be confirmed.}
\tablefoottext{b}{New classification.}
}
\end{table*}

\subsubsection{[SPH11]  472: XMMM31~J004047.02+405146.7}\label{472}

[SPH11] 472 is a soft X-ray source with hardness ratios
indicative of an SNR. As no foreground
star was found, it was suggested as an SNR candidate by PFH05 and SPH11.

No diffuse optical line emission was detected in the LGGS \ha, [\sii],
and [\oiii] images (Fig.\,\ref{lggs}2).
The LGGS $B$, $V$, $R$, and $I$ band images show a red optical 
point-like source inside the error circle of the \xmm\ source
(Fig.\,\ref{rband}, top right).
This optical source LGGS\,J004046.96+405148.2 has the colour 
$V - R$ = 1.4 and \logfxfo\ = --0.6, thus didn't only marginally
fulfill the criterion for stars of $V - R >$ 1, 
\logfxfo\ $ < -0.65$ (see Sect.\,9.1 in SPH11).
Therefore, we re-classify the source as a foreground star candidate.

\subsubsection{[SPH11]  500: XMMM31~J004052.88+403624.4}

[SPH11] 500 is a hard X-ray source.
Owing to the positional coincidence
with the optical SNR candidate with low confidence
suggested by MPV95 (source 3-027), the source was classified as 
an SNR candidate by SPH11. 

However, it is also listed in PFH05 and SBK09 as a hard source. 
An extended, diffuse source is visible at its position on the LGGS 
\ha\ image (Fig.\,\ref{lggs}2), however the flux ratio of [\sii]/\ha\ = 
$0.3\pm0.2$ is lower than what is typical for SNRs.
Therefore, this hard X-ray source with an optical counterpart that is
probably no SNR, is no longer an SNR candidate.

\subsubsection{[SPH11]  509: XMMM31~J004055.02+411216.4}

[SPH11] 509 is a soft X-ray source and was classified as an SNR candidate 
based on its hardness ratios by SPH11.

However, on the LGGS $B$, $V$, $R$, and $I$ band images, we find
the star LGGS\,J004054.86+411217.1 at the X-ray position with $V - R$ = 1.3 
and \logfxfo\ = --0.8 (Fig.\,\ref{rband}, upper middle left).
Therefore, we re-classify this sources as a foreground star candidate.

\subsubsection{[SPH11]  682: XMMM31~J004140.28+405947.9}

[SPH11] 682 is the faintest source in our list with a
detection likelihood of $ML = 7.6$. 
It was classified as an SNR candidate by SPH11
based on the hardness ratio criterion and the lack of a likely foreground
star at its position.

However, with $HR_2 = -0.48\pm0.26$, the criterion $HR_2 + EHR_2 < -0.2 $ 
is only marginally fulfilled. Furthermore, the fact that even $HR_4$ was 
determined indicates that there is a hard component.
Only faint \ha\ and [\sii] emission and no [\oiii] emission is detected
in the LGGS data (Table \ref{nosnrlist}). 
No radio counterpart is known either.
Therefore, we revise its classification and leave the 
source without classification.

\subsubsection{[SPH11] 1156: XMMM31~J004310.43+413850.1}\label{1156}

[SPH11] 1156 is a hard X-ray source,
which was classified as an SNR candidate
by SPH11 based on the positional coincidence with
the source K89A in BW93. 

[BW93] K89A is seen on the LGGS \ha\ and [\sii] images to the west of
the X-ray source position (Fig.\,\ref{lggs}5). It has a very faint 
[\oiii] emission and a flux ratio of [\sii]/\ha\ $\approx$ 1 in the 
north. Therefore, the optical source was discussed as an SNR candidate.
However, the \xmm\ source is located east of [BW93] K89A and does not 
seem to be related to the optical source.

New optical follow-up observations (Hatzidimitriou et al., in prep.) 
have shown that the spectrum taken at the position of the \xmm\ source
is a composite spectrum of an early-type star and a nebular source with
[\sii]/\ha\ = $0.57\pm0.09$. This is in agreement with the value
derived from the LGGS images (Table \ref{nosnrlist}).

Neither the X-ray emission nor the optical emission is thus
clearly indicative of an SNR at the position of [SPH11] 1156. 
Therefore, we classify the X-ray 
source as $<$hard$>$ based on the X-ray hardness ratios. 

\subsubsection{[SPH11] 1437: XMMM31~J004421.18+421621.9}

[SPH11] 1437 is a faint soft X-ray source with low $ML = 8.5$, which
was suggested as an SNR candidate based on its hardness ratios 
by SPH11 (see Fig.\,\ref{nolggs}).

The location of this source was not covered by the LGGS. Two faint sources 
are found on the SDSS-III $r$, $i$, and $z$ band images inside the \xmm\ error 
circle (Fig.\,\ref{sdss1437}). 
These sources are not visible in the $u$ band image.  
The sources are listed in the USNO-B1.0 catalogue 
\citep{2003AJ....125..984M}, both with
$R$ = 20.6, 21.4 and $B$ = 19.6, 19.1, respectively.
The \xmm\ source [SPH11] 1437 is therefore most probably either a star
or an AGN.
We therefore no longer classify this source as an SNR candidate.

\subsubsection{[SPH11] 1505: XMMM31~J004442.73+415340.5}\label{1505}

\begin{figure}
\centering     
\includegraphics[width=0.45\textwidth,clip=]{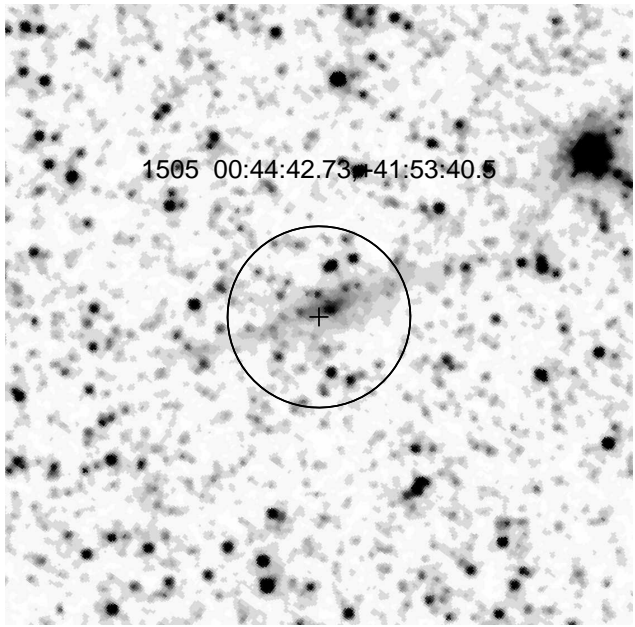}
\caption{
\hubble\ Space Telescope ACS/WFC (F814W) image showing
an extended source at the position of [SPH11] 1505, which seems to 
be a galaxy with an apparent extent of $\sim$6\arcsec\ $\times$ 1\arcsec.
It obviously has a nucleus located $\sim$0\farcs5 distant from the 
\xmm\ source position.
The shown area has a size of $\sim$15\arcsec\ $\times$ 15\arcsec. 
}
\label{hst1505}
\end{figure}

[SPH11] 1505 is a hard X-ray source with hardness ratios of 
$HR_1 = 0.50\pm0.27$, $HR_2 = -0.73\pm0.27$, $HR_3 = 0.97\pm0.04$, 
and $HR_4 =  0.50\pm0.04$. It was classified as an SNR candidate 
by SPH11 based on its hardness ratios $HR_1$ and $HR_2$ and the
lack of a foreground star at its position.

The source is located south of the \hii\ region [PAV78] 799 
(Fig.\,\ref{lggs}9). It is obviously not related to the \hii\ region.
Instead, there is apparently a galaxy at its position 
as can be seen on an \hst\ image taken with ACS/WFC using the filter 
F814W (Fig.\,\ref{hst1505}). This galaxy seems to have a clearly
visible nucleus, which is located inside the 3$\sigma$ error circle of 
the \xmm\ source position. 
This source also has an extended infrared counterpart in the Wide-field 
Infrared Survey Explorer (WISE) All-sky data release 
\citep{2012yCat.2311....0C}.
Therefore, we classify [SPH11] 1505 as an AGN candidate.

\subsubsection{[SPH11] 1534: XMMM31~J004450.02+415240.9}

[SPH11] 1534 is a soft X-ray source
and was classified as an SNR candidate by SPH11 based on its hardness ratios.

It is located near the optical SNR [BW93] K567 and its radio counterpart 
[B90] 308, which is seen at the northern edge of the \ha\ image 
(Fig.\,\ref{lggs}9). 
However, two faint red sources are found on, e.g., the LGGS
$R$ band image at the position of the \xmm\ source, which are
most likely stars (Fig.\,\ref{rband}, lower middle left).
The X-ray source seems not to be related to the SNR to the north.
Because optical point sources are found, we no longer classify the X-ray
source as an SNR candidate.

\subsubsection{[SPH11] 1608: XMMM31~J004517.87+420502.1}

[SPH11] 1608 is a faint soft X-ray source,
which was classified as an SNR candidate by SPH11 
based on its hardness ratios.

In the LGGS data, only faint \ha\ emission was detected. 
The LGGS $R$, $V$, and $B$ band images show a point source 
at the position of the \xmm\ source (Fig.\,\ref{rband}, lower middle right).
This optical source (LGGS\,J004517.84+420501.7) has $V - R$ = 1.2 and 
\logfxfo\ = --0.6, thus did not marginally fulfill the criterion for stars 
(Sect.\,9.1 in SPH11, see also Sect.\,\ref{472}).
However, as there is a star within the X-ray error circle and 
no additional extended source in the optical that is indicative
of an SNR, we now classify the X-ray source as a foreground star candidate.

\subsubsection{[SPH11] 1637: XMMM31~J004528.35+414605.9}

[SPH11] 1637 is a hard X-ray source.
It was suggested to be an SNR candidate by PFH05 and SPH11 as it is 
located near the SNR candidate [MPV95] 1-013.

The source [MPV95] 1-013 can be found at the top edge of the LGGS
\ha\ and [\sii] images in Fig.\,\ref{lggs}11. The X-ray source is located 
south of it and does not seem to be related to the SNR candidate.
Therefore, we no longer classify this hard X-ray source as an SNR candidate.

\subsubsection{[SPH11] 1712: XMMM31~J004556.01+421117.2}\label{1712}

[SPH11] 1712 is a faint, hard X-ray source.
It is coincident with the optical source [BW93] K884 and was therefore
suggested as an SNR candidate by SPH11.

The optical source [BW93] K884, however, has been suggested to be a 
supershell (Fig.\,\ref{lggs}11). 
The [\sii]/\ha\ ratio derived from the LGGS data is 0.5$\pm$0.2
and hence more consistent with an \hii\ region.

The optical spectrum obtained in a new follow-up observation
by Hatzidimitriou et al.\ (in prep.) indicates
a composite of a most likely highly reddened Be star and 
nebular emission.
The [\sii]/\ha\ ratio of 0.36$\pm$0.08 is consistent with the LGGS value
and not indicative of an SNR.

As the X-ray source is hard and there is no optical emission suggesting
an SNR nature, we no longer classify this source as an SNR candidate. 

\subsubsection{[SPH11] 1732: XMMM31~J004602.53+414513.2}

[SPH11] 1732 is a hard X-ray source and
was classified as an SNR candidate by PFH05 and SPH11 
as it is located near the source [MPV95] 3-111, which had been 
suggested to be an SNR candidate.

However, [SPH11] 1732 is located $\sim$30\arcsec\ distant from [MPV95] 3-111.
It is located at the edge of a large diffuse structure seen in \ha\ 
(Fig.\,\ref{lggs}11), but shows no clear correlation with the optical line
emission. 
Therefore, we classify this X-ray source as a hard source and
not as an SNR candidate.

\subsubsection{[SPH11] 1741: XMMM31~J004604.42+414943.2}

[SPH11] 1741 is a soft X-ray source 
with no optical counterpart that seems to be a foreground star.
Therefore, the source was classified as an SNR candidate 
by PFH05 and SPH11.
It has also been detected with \chandra\ ([WGK04], [DKG04] n1-48). 

We find two optical point sources on the LGGS $B$, $V$, and $R$ band 
images, one inside and one right outside the error circle of the \xmm\ 
source (Fig.\,\ref{rband}, bottom). 
The optical source inside the error circle (LGGS\,J00460.38+414944.0) 
has the optical colour $V - R = 1.3 > 1$ and \logfxfo\ = $-1.3 < -0.65$, 
indicative of a star.
Therefore, we classify the X-ray source as a foreground star candidate.

\section{Comparison to other X-ray catalogues}

\subsection{Comparison to the \rosat\ catalogue}



In the \rosat\ source catalogues of SHP97 and SHL01, 16 sources were
classified as SNRs. Thirteen out of the 16 sources were 
detected in the \xmm\ survey.
No counterpart was found for the \rosat\ sources [SHL01] 129 
(RX~J0041.9+4046), 
[SHP97] 203 (RX~J0042.8+4125), 
and [SHP97] 258 (RX~J0043.9+4152).

The \rosat\ source [SHP97] 284 (RX~J0044.6+4125) is located 
between two \xmm\ sources: [SPH11] 1497, which is an SNR with optical and 
radio counterparts, and 1481, which was classified as an SNR candidate
by SPH11 based on its X-ray hardness ratios.
Based on the new optical data of the LGGS, we confirmed [SPH11] 1481
as an SNR.


We have classified [SPH11] 1079 ([SHP97] 212) as a new 
SNR based on the optical line emission. 
In addition, we have confirmed that three \xmm\ sources 
[SPH11] 472, 682, and 1637
with \rosat\ counterparts
([SHP97] 92, [SHL01] 110, and [SHL01] 321, respectively)
are no SNR candidates.

\subsection{Comparison to \chandra\ sources}\label{compchan}

\xmm\ detected five of six SNRs in M\,31 that were classified based on 
\chandra\ observations:
[KGP02] r2-57 ([SPH11] 883), 
[KGP02] r3-84 ([SPH11] 1040), 
[KGP02] r2-56 ([SPH11] 1050), 
[KGP02] r3-69 ([SPH11] 1060), and
[KGP02] r3-63 ([SPH11] 1234). 
The \chandra\ source CXOM31\,J004247.82+411525.7 that was identified
as an SNR by Kong et al.\ (\citeyear{2003ApJ...590L..21K}, source 2) is 
located between two bright sources and has not been resolved by \xmm.

\citet{2004ApJ...609..735W} classified the \chandra\ source
[WGK04] n1-85 (also [DKG04]) as a variable X-ray source coincident with a
radio SNR ([B90] 265). In addition, this source was identified as a 
transient source by \citet{2006ApJ...643..356W}. 
It was not detected by \xmm\ even 
though the position is covered by \xmm\ observations, confirming
its transient nature. Therefore, this source is not likely an SNR.

\section{Discussion}

\subsection{Cumulative X-ray luminosity distribution}\label{cumxlf}

\begin{figure}
\centering
\includegraphics[width=0.5\textwidth,bb=50 0 440 355,clip=]{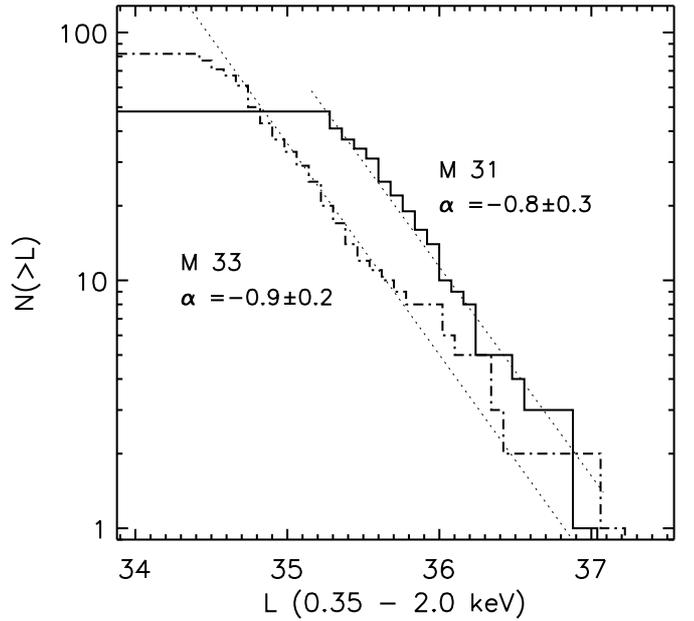} 
\caption{
Cumulative luminosity distribution of X-ray SNRs and candidates 
in M\,31 listed in Tables \ref{snrlist} and \ref{candlist}
for the band of 0.35 -- 2.0~keV (solid line).
For comparison, the distribution of SNRs in M\,33 is also shown 
\citep[dash-dotted,][]{2010ApJS..187..495L}.
The dotted lines show the power-law fits. 
The power-law index $\alpha$ is given.
}
\label{xlf}
\end{figure}

We estimated the luminosity of the SNRs and candidates listed
in Tables \ref{snrlist} and \ref{candlist} and thus obtained a cumulative
luminosity distribution of the SNRs in M\,31. 
For the M\,31 SNRs we converted the \xmm\ count rates into flux by assuming a 
thermal spectrum with $kT = 0.2$~keV, absorbed by a foreground column density of 
\nh(MW) $= 0.7 \times 10^{21}$~cm$^{-2}$ and an additional 
\nh(M31) $= 1.0 \times 10^{21}$~cm$^{-2}$ similar to values of the fits of the 
four brightest sources [SPH11] 969, 1050, 1066, and 1234. 
We simulated the flux assuming different models by varying the
temperature in the range of $kT = 0.1 - 0.3$~keV and the foreground 
absorption \nh(M31) $= 0 - 2 \times 10^{21}$~cm$^{-2}$ for the observed
count rates and obtained an uncertainty of $\sim$20\%.
Assuming CIE or NEI model results in a difference in flux of $\sim$10\%
(see Table \ref{10661234}). As most of the sources are too faint to 
distinguish between CIE and NEI, we assume the same {\tt APEC} model
for all sources.
For the cumulative luminosity distribution plot shown in Fig.\,\ref{xlf}, the 
luminosities in the energy band of 0.35 -- 2.0~keV were calculated with the 
foreground absorption set to zero.
We also included the source CXOM31\,J004247.82+411525.7, which was identified
as an SNR by \citet{2003ApJ...590L..21K} but not resolved with \xmm\ (see
Sect.\,\ref{compchan}).
For comparison, the luminosities for M\,33 SNRs taken from 
\citet{2010ApJS..187..495L} are plotted as well. These luminosities were
converted from \chandra\ count rates by \citet{2010ApJS..187..495L} assuming 
a thermal plasma model with $kT = 0.6$~keV and an absorbing 
\nh\ $= 5.0 \times 10^{20}$~cm$^{-2}$, which correspond to the best fit
values for the brightest SNRs in M\,33 observed with \chandra.

As one can see in Fig.\,\ref{xlf} the slope of the cumulative luminosity 
distribution of X-ray SNRs in M\,31 and M\,33 are comparable. Both 
distributions can be fitted with a power law with an index of 
$\alpha \approx -1$. 
The distribution
in M\,33 seems to deviate from this power-law distribution for luminosities
$> 5 \times 10^{35}$~erg~s$^{-1}$ showing an excess.
The slight difference in the shape of the cumulative luminosity distribution 
might indicate that the fraction of more luminous SNRs in M\,33 is higher than 
in M\,31. 
There are 24 sources brighter than $5 \times 10^{35}$~erg~s$^{-1}$
in M\,31 and nine in M\,33. Above $10^{36}$~erg~s$^{-1}$, there are
13 sources in M\,31 and seven in M\,33 
\citep[see also][]{2005AJ....130..539G,2010ApJS..187..495L}.

The number of SNRs in M\,33 is higher than what is expected if we simply
scale with the total mass of the galaxy, as M\,33 is about 10 times less 
massive 
than M\,31 \citep{2003MNRAS.342..199C,2010MNRAS.406..264W}.
M\,33 is a typical flocculent spiral galaxy with 
discontinuous spiral arms, in which star formation regions are found. 
In contrast to grand design spirals such as the Milky Way or M\,31, in which 
density waves are 
believed to produce the spiral arms,
gravitational instabilities together with turbulence in the interstellar medium
seem to be the origin of the spiral structure and thus the on-going star formation 
in flocculent galaxies 
\citep[and references therein]{1984ApJ...282...61S,2003ApJ...593..333E}.
The star formation rate in the disk of M\,31 is 0.27~$M_{\sun}$~yr$^{-1}$
for $6 < R < 17$~kpc
\citep{2010A&A...517A..77T} corresponding to a star formation rate per unit
area of $\Sigma_{\rm SFR} = 0.4 M_{\sun}$~Gyr$^{-1}$~pc$^{-2}$.
This value is about six times lower than in M\,33, 
for which a star formation rate per unit area of
$\Sigma_{\rm SFR} = 2 - 3 M_{\sun}$~Gyr$^{-1}$~pc$^{-2}$ 
has been measured \citep{2009A&A...493..453V}.
The higher star formation rate implies a higher rate for the occurrence of 
core-collapse SNRs in M\,33.

In addition, one should note that our list of SNRs in M\,31 is based on \xmm\
observations whereas the M\,33 SNRs have been detected in a survey performed 
with \chandra. Not only was the \chandra\ survey of M\,33 deeper, but the 
superior angular resolution of \chandra\ made it possible to detect smaller 
and thus most likely younger SNRs in M\,33, which would not have been 
resolved and classified 
as an SNR in an \xmm\ observation. However, the \chandra\ M\,33 survey only 
observed the inner part of the galaxy inside the $D_{25}$ ellipse, whereas the \xmm\
M\,31 survey fully covered the $D_{25}$ ellipse.

\subsection{Radial distribution}

\begin{table*}
\caption{
\label{info}
Geometric parameters of M\,31 and M\,33
used for the calculation of the radial SNR number density distribution.
}
\centering
\begin{tabular}{cccccc}
\hline\hline
Galaxy & Distance\tablefootmark{a} [kpc] & Position Angle\tablefootmark{b} & Inclination Angle\tablefootmark{b} & Corrected $D_{25}$\tablefootmark{b} & $R_{25}$ [kpc] \\
\hline
M\,31 & 744 & 35\degr\ & 71\degr\ & 155\farcm5 & 16.8 \\
M\,33 & 805 & 23\degr\ & 54\degr\ & 52\farcm6 & 6.2 \\
\hline
\end{tabular}
\tablefoot{
\tablefoottext{a}{\citet{2010A&A...509A..70V} for M\,31 and 
\citet{2009MNRAS.396.1287S} for M\,33.}
\tablefoottext{b}{\citet{1991trcb.book.....D}.}
}
\end{table*}

In the shock waves of SNRs, particles can gain energies up to 10$^{15}$~eV or
higher due to diffusive shock acceleration \citep[][]
{1978MNRAS.182..147B,1978MNRAS.182..443B,1978ApJ...221L..29B,2005JPhG...31R..95H}.
Therefore, SNRs together with pulsars are
thought to be the primary sources of Galactic cosmic rays (CRs).
The distribution of SNRs and pulsars in our Galaxy is a crucial basis for the
understanding of the CR distribution. \citet{1977ApJ...217..843S} studied the
radial dependence of the Galactic SNR surface density using
observational SNR data of \citet{1974PASJ...26..255K}
and pulsar data of \citet{1977NASCP.002..265S}
and showed that it can be described as $\propto x^{\alpha} \exp{(-x/R)}$
with $x$ being the radial distance to the Galactic centre.
Using only Galactic shell-type SNRs, \citet{1989PASP..101..607L}
found a peak in the surface density distribution at 4 -- 6~kpc
distance from the Galactic centre.
\citet{1998ApJ...504..761C} re-analysed the Galactic SNR data and
suggested to use a dependence of the type $\propto sin(\pi x + \theta)
\exp{(-x/R)}$ for the radial surface density distribution. They
obtained a scale length of $\sim7$~kpc and a maximum of the distribution
at about 5~kpc.
Based on the obtained distribution of SNRs in our Galaxy, the spectral 
distribution of Galactic CRs can be modelled to explain the CR spectrum
up to 10$^{15}$ -- 10$^{16}$~eV \citep[see, e.g.,][]{2005JPhG...31R..95H}.
However, our vantage point is not ideal to study the source
distribution in our Galaxy. To understand the distribution of SNRs in a 
spiral galaxy, it is thus necessary to study the most nearby spiral galaxies 
M\,31 and M\,33.

\begin{figure}
\centering
\includegraphics[width=0.4\textwidth,bb=63 0 435 354,clip=]{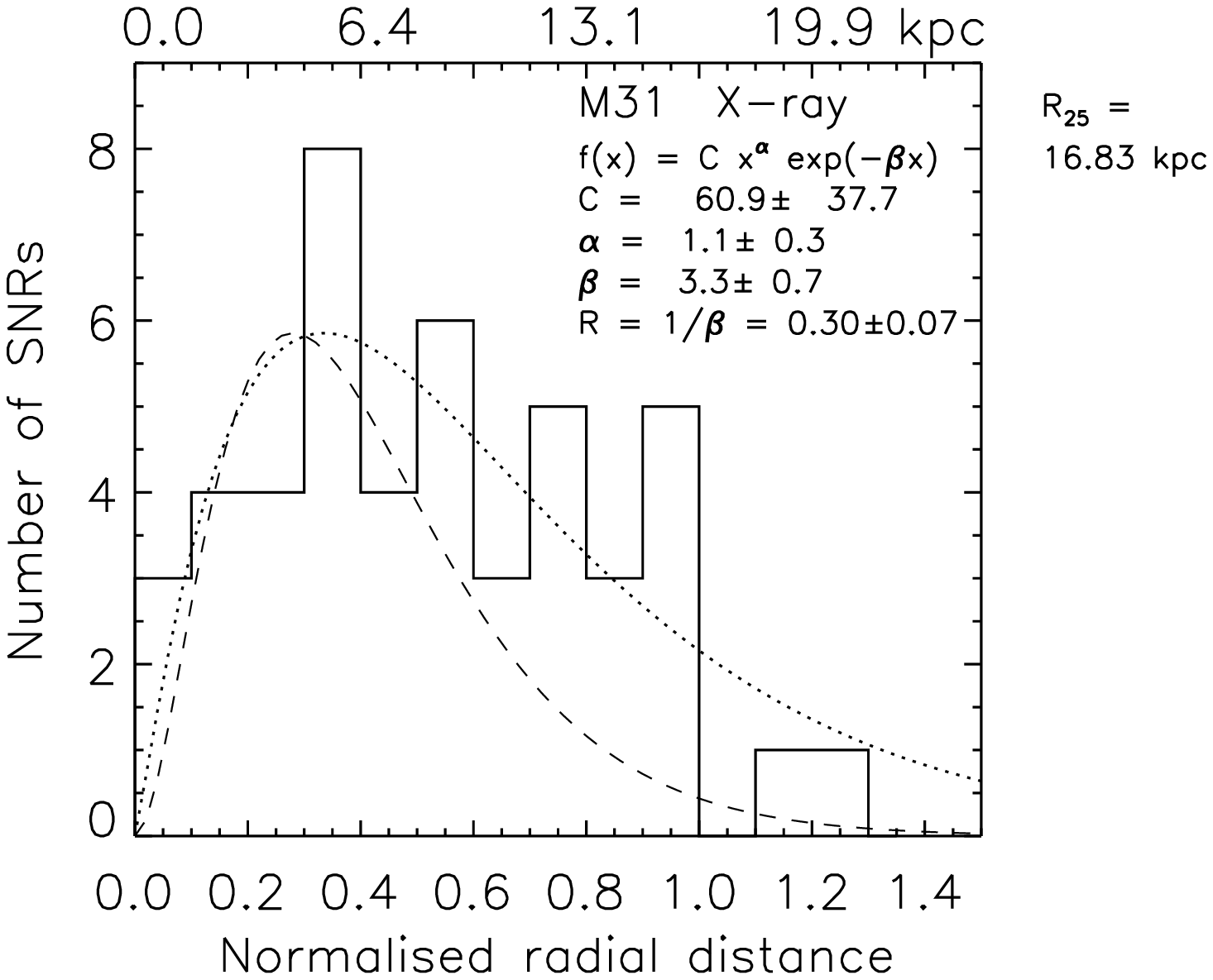} \\[2mm]
\includegraphics[width=0.4\textwidth,bb=63 0 435 354,clip=]{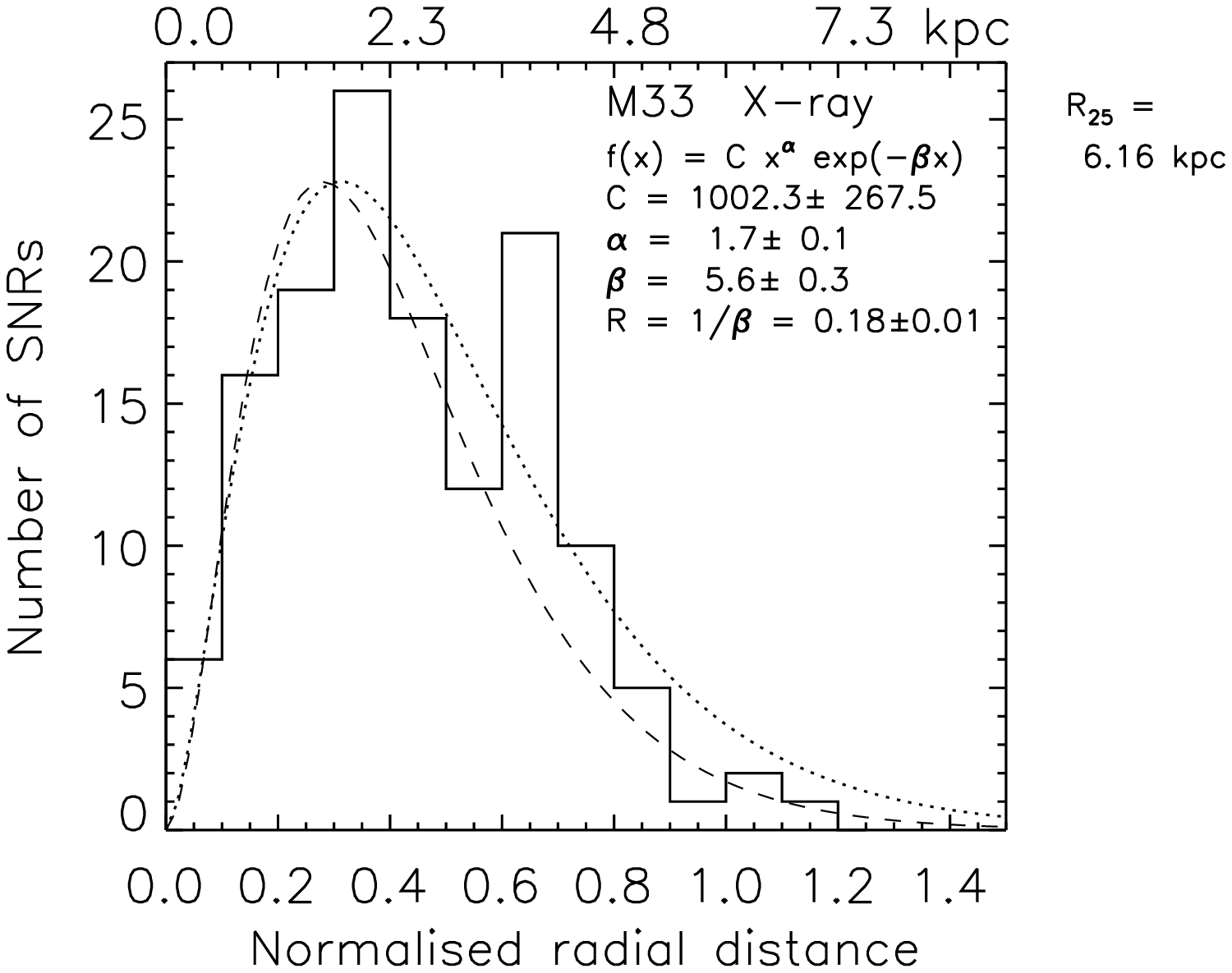} 
\caption{
Surface density of SNRs and candidates in M\,31 (this work) and M\,33
\citep{2010ApJS..187..495L} plotted over the radius normalised to $D_{25}/2$
\citep[$D_{25}$ = 155.5\arcmin, 52.6\arcmin\ for M\,31 and M\,33, 
respectively,][]{1991trcb.book.....D}. The radial distance in kpc is given
along the upper x-axis.
Dotted lines show the fitted model function
$f(x) = C x^{\alpha} \exp{(-\beta x)}$, 
dashed lines show the model function for the Milky Way normalised to
the maximum of M\,31 or M\,33.
}
\label{radial}
\end{figure}

In order to obtain the radial distribution of SNRs in M\,31,
the positions of the SNRs and candidates in Tables \ref{snrlist}
and \ref{candlist} as well as the source CXOM31\,J004247.82+411525.7 
were first corrected for projection and their
galactocentric distances were computed.
\citet{2011A&A...534A..55S} have also presented the radial distribution 
of SNRs and candidates in M\,31 detected in the \xmm\ LP.
For comparing M\,31 and M\,33,
the radial distances were normalised to $R_{25} = D_{25}/2$. 
The positions of sources were binned into equidistant radial bins and 
the surface density was calculated for each annulus. 
The parameters used, i.e., distance, inclination angle, position angle, and 
$D_{25}$ are listed in Table \ref{info}.
The SNR surface densities are plotted against the normalised radial
distance in Fig.\,\ref{radial}.
We fitted the obtained radial surface density distribution of the
SNRs in each galaxy with the model distribution introduced for the
radio selected SNRs in the Milky Way
\begin{equation}\label{xexp}
f(x) = C\,x^{\alpha} \exp{(-x/R)}
\end{equation}
with a maximum at a radial distance of several kpc from the Galactic centre
as originally suggested by \citet{1977ApJ...217..843S}.
The parameters of the distribution obtained by \citet{1998ApJ...504..761C}
for our Galaxy are $C = 136.5, \alpha = 2.00$, and $R = 0.14$.
The best-fit model curves according to Eq.\,\ref{xexp} are plotted in
Fig.\,\ref{radial} with dotted lines. 
This radial SNR distribution in M\,31 and M\,33 is different from the mass 
distribution of these galaxies derived from the rotation curves 
\citep[e.g.,][and references therein]{2003MNRAS.342..199C,2010A&A...511A..89C}.
The SNR distribution rather seems to follow the distribution of stars and 
gas, similar to what had been suggested for the Milky Way. 
For comparison, we also plot the distribution for the Milky Way
normalised to the fitted maxima of the distributions in M\,31 and M\,33.
The fitted curve indicates
a maximum at about 5.5~kpc and 2~kpc for M\,31 and M\,33, respectively,
corresponding to $\sim$0.3~$R_{25}$ for both galaxies.
The distribution in M\,31 seems to be almost flat for
$<$ 17~kpc $\approx R_{25}$ and falls exponentially outside 
$\sim1.0~R_{25}$, while the distribution in M\,33 falls off exponentially
for $>$ 4~kpc $\approx 0.65~R_{25}$. 
This behaviour of the SNR distribution in M\,31 seems to be correlated with 
the distribution of gas in M\,31, which is known to have ring-like structures 
consisting of many spiral arms between a radius of $\sim5$~kpc to 20~kpc, 
with the most prominent ring found at a radius of $\sim10$~kpc 
\citep{1981PASJ...33..449S,1991ApJ...372...54B}.

\section{Summary}

We have studied the X-ray and optical properties of SNRs and candidates 
in the source catalogue of the \xmm\ LP survey of M\,31.
For the optical analysis we used the \ha, [\sii], and [\oiii] emission line
images as well as the $UBVRI$ band images of the LGGS 
\citep{2006AJ....131.2478M}, which covered M\,31 in ten fields.

In addition to the 56 SNRs and candidates in the \xmm\ survey source catalogue 
by SPH11, we found seven sources, which had not been classified as
SNR candidates but showed optical line emission indicative of SNRs. 
Therefore, we included these seven sources also in the list of studied 
sources. We extracted X-ray spectra of the twelve brightest 
sources. Only four of the sources have high enough statistics to perform
a more detailed analysis of the X-ray spectra taken with the \xmm\ EPIC
instruments. 
For the three brightest sources ([SPH11] 969, 1066, 1234) a collisional
ionisation equilibrium model ({\tt APEC}) does not reproduce the spectra well, 
while a non-equilibrium ionisation model ({\tt NEI}) improves the fit.
Two thermal components with two different temperatures are necessary for 
the spectrum of [SPH11] 1234 to achieve a satisfactory fit indicating that
the X-ray emission is a superposition of at least two emission components.

For each SNR and candidate we measured the \ha, [\sii], and [\oiii] fluxes
as well as the  [\sii]/\ha\ flux ratio and thus confirmed five new
X-ray sources as SNRs. 
We want to point out that the final classification of this work applies 
to the X-ray source. Therefore, if, for example,
the optical counterpart is likely an SNR
but the X-ray source is either a soft source with a star in the
positional error circle or is a hard source, the X-ray source was classified 
as an SNR candidate.

We identified 17 sources, which are either hard X-ray sources with no
optical counterpart indicative of an SNR or with
an optical counterpart that is likely a star or a background AGN. Therefore, 
we excluded these sources from the list of SNRs or candidates. 
In particular, source [SPH11] 1505, for which we found a galaxy with a bright 
nucleus as an optical counterpart, has been classified as an AGN candidate.

From the \xmm\ LP survey catalogue, we have thus obtained a list of 
26 X-ray SNRs and additional 20 bona-fide X-ray SNR candidates in M\,31.
The brightest SNRs have X-ray luminosities of $\sim8 \times 10^{36}$~erg~s$^{-1}$ 
in the 0.3 -- 2.0~keV band. 

\begin{acknowledgements}
We have made extensive use of the optical data from the Local Group Galaxy Survey,
which are kindly made available by Phil Massey and his collaborators.
\\
This work has also made use of {\it SkyView} and images from the Sloan Digital Sky
Survey (SDSS) III.
Funding for SDSS-III has been provided by the Alfred P.\ Sloan Foundation, the 
Participating Institutions, the National Science Foundation, and the U.S.\ Department 
of Energy. The SDSS-III web site is http://www.sdss3.org/.
SDSS-III is managed by the Astrophysical Research Consortium for the Participating 
Institutions of the SDSS-III Collaboration including the University of Arizona, the 
Brazilian Participation Group, Brookhaven National Laboratory, University of Cambridge, 
University of Florida, the French Participation Group, the German Participation Group, 
the Instituto de Astrofisica de Canarias, the Michigan State/Notre Dame/JINA Participation 
Group, Johns Hopkins University, Lawrence Berkeley National Laboratory, Max Planck 
Institute for Astrophysics, New Mexico State University, New York University, Ohio 
State University, Pennsylvania State University, University of Portsmouth, Princeton 
University, the Spanish Participation Group, University of Tokyo, University of Utah, 
Vanderbilt University, University of Virginia, University of Washington, and Yale 
University.
\\
Furthermore, this work is based on observations made with the NASA/ESA \hubble\ Space 
Telescope, obtained from the data archive at the Space Telescope Institute. STScI is 
operated by the association of Universities for Research in Astronomy, Inc.\ under the 
NASA contract NAS 5-26555. 
\\
This publication makes use of data products from the Wide-field Infrared Survey
Explorer, which is a joint project of the University of California, Los Angeles, and
the Jet Propulsion Laboratory/California Institute of Technology, funded by the
National Aeronautics and Space Administration.
\\
M.S.\ acknowledges support by the Deutsche Forschungsgemeinschaft through the Emmy 
Noether Research Grant SA 2131/1.
\end{acknowledgements}

\bibliographystyle{aa} 
\bibliography{../../bibtex/ctb109,../../bibtex/xraytel,../../bibtex/my,../../bibtex/cosmicray,../../bibtex/ism,../../bibtex/nearbygal}



\Online

\begin{appendix}

\onecolumn

\section{\xmm\ and LGGS images}\label{lggs}

{\bf See 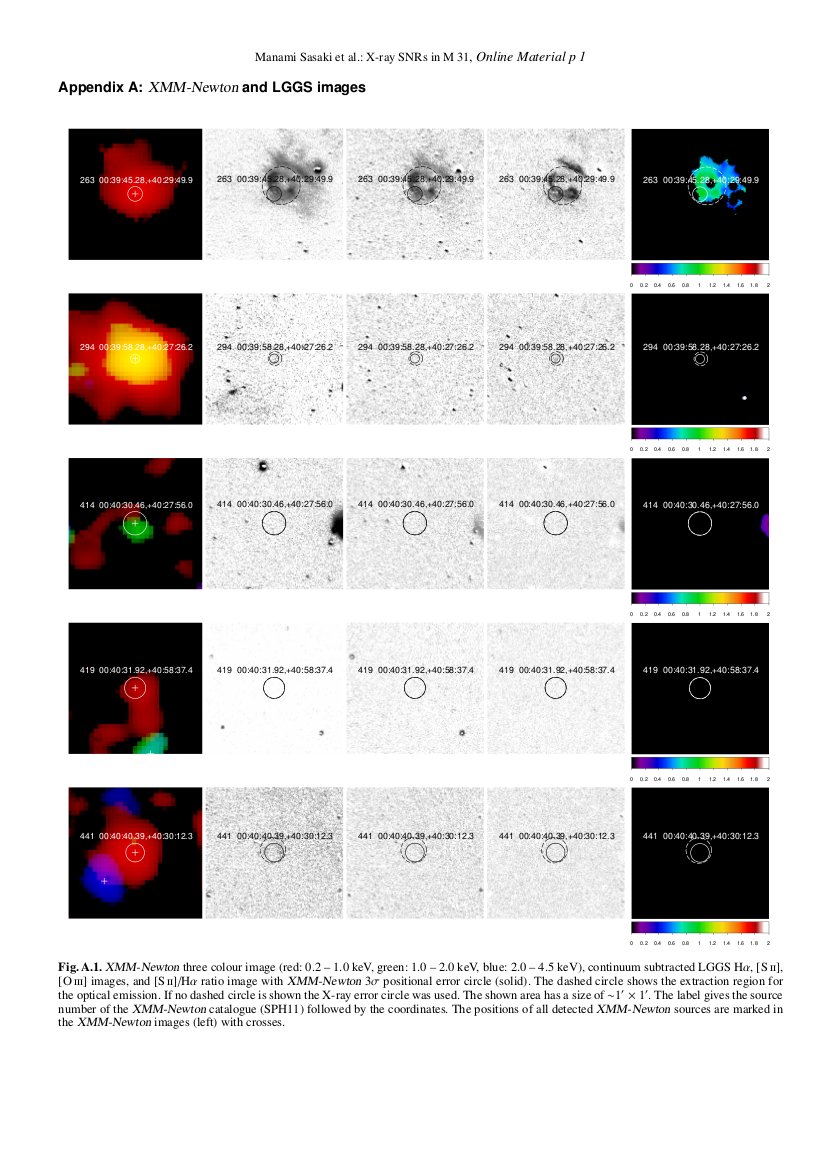 to 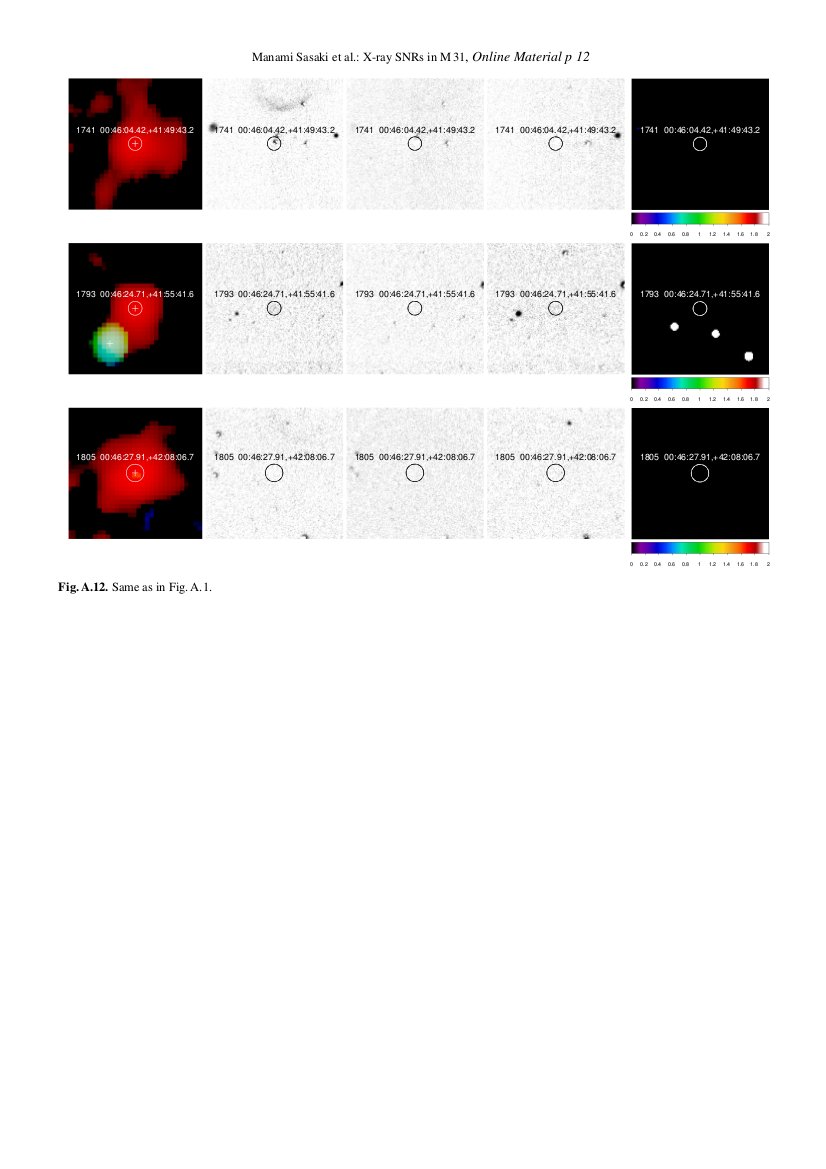.}

\end{appendix}

\clearpage

\begin{appendix}

\section{Supernova remnants and candidates in the \xmm\ M\,31 large survey
catalogue}

\begin{table*}[ht]
\caption{
\label{xlist}
X-ray SNRs and candidates in the \xmm\ M\,31 survey
catalogue by \citet{2011A&A...534A..55S}.
}
\centering
\begin{tabular}{rccccrcccc} 
\hline\hline
[SPH11] & RA (2000.0) & Dec (2000.0) & Pos.\ error & Rate & $ML$\tablefootmark{a} & $HR_1$\tablefootmark{b} & $HR_2$\tablefootmark{b} & $HR_3$\tablefootmark{b} & $HR_4$\tablefootmark{b} \\
\multicolumn{1}{c}{ID} &  &  & [arcsec] & [cts/s] & & & & \\ 
\hline
\multicolumn{10}{c}{SNRs\tablefootmark{c}} \\
\hline
 182 & 00 39 23.50 & +40 44 19.1 & 5.30   & 2.0e--3$\pm$5.0e--5 &  13 & 0.95$\pm$0.27 & --0.64$\pm$0.20 &                 &                 \\
 263 & 00 39 45.28 & +40 29 49.9 & 3.37   & 3.7e--3$\pm$4.0e--5 &  96 & 0.42$\pm$0.10 & --0.93$\pm$0.09 &                 &                 \\
 474 & 00 40 47.19 & +40 55 24.6 & 2.69   & 6.4e--3$\pm$6.0e--5 & 180 & 0.90$\pm$0.06 & --0.63$\pm$0.08 & --0.88$\pm$0.20 &                 \\
 668 & 00 41 35.76 & +41 06 55.0 & 4.50   & 9.8e--3$\pm$2.0e--4 &  62 & 0.71$\pm$0.17 & --0.30$\pm$0.15 & --0.80$\pm$0.28 &   0.83$\pm$0.25 \\
 883 & 00 42 24.41 & +41 17 29.8 & 2.65   & 3.7e--3$\pm$4.0e--5 &  52 & 0.80$\pm$0.07 & --0.98$\pm$0.06 &                 &                 \\ 
1040 & 00 42 49.11 & +41 24 06.6 & 2.38   & 9.2e--3$\pm$5.0e--5 & 330 & 0.48$\pm$0.05 & --0.84$\pm$0.05 & --1.00$\pm$0.27 &                 \\ 
1050 & 00 42 50.47 & +41 15 56.7 & 3.28   & 1.1e--2$\pm$9.0e--5 &  73 & 0.53$\pm$0.08 & --0.96$\pm$0.06 &                 &                 \\ 
1066 & 00 42 53.53 & +41 25 51.0 & 1.89   & 3.5e--2$\pm$1.0e--4 &1400 & 0.56$\pm$0.03 & --0.70$\pm$0.03 & --0.94$\pm$0.11 &                 \\ 
1234 & 00 43 27.93 & +41 18 30.5 & 1.79   & 7.4e--2$\pm$1.0e--4 &6300 & 0.44$\pm$0.02 & --0.71$\pm$0.01 & --0.95$\pm$0.03 &                 \\ 
1275 & 00 43 39.27 & +41 26 53.0 & 2.12   & 2.9e--2$\pm$1.0e--4 &1200 & 0.39$\pm$0.04 & --0.76$\pm$0.04 & --0.87$\pm$0.12 &                 \\ 
1291 & 00 43 43.97 & +41 12 32.7 & 2.15   & 1.7e--2$\pm$8.0e--5 & 620 & 0.77$\pm$0.04 & --0.45$\pm$0.04 & --0.94$\pm$0.06 &                 \\ 
1328 & 00 43 53.69 & +41 12 04.4 & 2.80   & 1.1e--2$\pm$8.0e--5 & 270 & 0.53$\pm$0.07 & --0.58$\pm$0.07 & --1.00$\pm$0.12 &                 \\ 
1351 & 00 43 58.25 & +41 13 28.7 & 3.17   & 6.0e--3$\pm$6.0e--5 &  86 & 0.42$\pm$0.10 & --0.78$\pm$0.10 & --0.48$\pm$0.43 &                 \\ 
1372 & 00 44 04.71 & +41 48 46.7 & 6.39   & 2.0e--3$\pm$6.0e--5 & 9.0 & 0.78$\pm$0.28 & --0.55$\pm$0.24 &                 &   1.00$\pm$0.26 \\
1386 & 00 44 09.53 & +41 33 20.9 & 4.49   & 2.0e--3$\pm$3.0e--5 &  26 & 0.42$\pm$0.17 & --0.78$\pm$0.13 &                 &                 \\ 
1410 & 00 44 13.55 & +41 19 54.3 & 3.26   & 7.8e--3$\pm$8.0e--5 & 120 & 0.92$\pm$0.07 & --0.37$\pm$0.09 & --0.72$\pm$0.17 &                 \\ 
1497 & 00 44 38.91 & +41 25 28.8 & 5.17   & 5.4e--3$\pm$1.0e--4 &  20 & 0.72$\pm$0.17 & --0.46$\pm$0.19 & --0.68$\pm$0.29 &                 \\ 
1522 & 00 44 47.19 & +41 29 18.7 & 4.83   & 4.1e--3$\pm$6.0e--5 &  34 & 0.44$\pm$0.14 & --0.70$\pm$0.14 &                 &                 \\ 
1535 & 00 44 51.07 & +41 29 05.9 & 2.71   & 1.5e--2$\pm$1.0e--4 & 250 & 0.18$\pm$0.08 & --0.22$\pm$0.09 & --0.51$\pm$0.12 & --0.55$\pm$0.53 \\ 
1539 & 00 44 52.82 & +41 54 58.1 & 3.81   & 1.9e--3$\pm$3.0e--5 &  22 & 0.70$\pm$0.21 & --0.08$\pm$0.18 & --0.62$\pm$0.23 &                 \\ 
1587 & 00 45 10.59 & +41 32 51.3 & 4.69   & 3.3e--3$\pm$7.0e--5 &  22 & 0.31$\pm$0.19 & --0.67$\pm$0.35 &                 &                 \\ 
1593 & 00 45 12.31 & +42 00 29.6 & 4.16   & 1.6e--3$\pm$3.0e--5 &  13 & 0.89$\pm$0.16 & --0.43$\pm$0.18 & --0.10$\pm$0.32 &                 \\ 
1599 & 00 45 13.94 & +41 36 15.5 & 2.52   & 1.9e--2$\pm$1.0e--4 & 590 & 0.71$\pm$0.05 & --0.55$\pm$0.05 & --1.00$\pm$0.05 &                 \\ 
1793 & 00 46 24.71 & +41 55 41.6 & 3.18   & 4.7e--3$\pm$5.0e--5 & 140 & 0.55$\pm$0.09 & --0.66$\pm$0.10 & --0.87$\pm$0.37 &                 \\ 
1805 & 00 46 27.91 & +42 08 06.7 & 4.03   & 7.4e--3$\pm$8.0e--5 & 120 & 0.45$\pm$0.09 & --0.73$\pm$0.10 & --1.00$\pm$0.43 &                 \\
\hline
\multicolumn{10}{c}{SNR candidates\tablefootmark{c}} \\
\hline
 294 & 00 39 58.28 & +40 27 26.2 & 2.08   & 1.9e--2$\pm$8.0e--5 &1300 & 0.64$\pm$0.03 & --0.61$\pm$0.04 & --0.95$\pm$0.07 &                 \\ 
 414 & 00 40 30.46 & +40 27 56.0 & 5.38   & 4.1e--3$\pm$8.0e--5 &  19 &               &   0.45$\pm$0.23 & --0.34$\pm$0.21 & --0.22$\pm$0.39 \\
 419 & 00 40 31.92 & +40 58 37.4 & 4.85   & 2.1e--3$\pm$4.0e--5 &  15 & 0.20$\pm$0.20 & --0.82$\pm$0.23 &                 &                 \\ 
 441 & 00 40 40.39 & +40 30 12.3 & 4.28   & 3.7e--3$\pm$6.0e--5 &  41 & 0.75$\pm$0.12 & --0.74$\pm$0.15 &                 &                 \\ 
 472 & 00 40 47.02 & +40 51 46.7 & 4.04   & 3.5e--3$\pm$5.0e--5 &  44 & 0.61$\pm$0.13 & --0.70$\pm$0.15 &                 &                 \\ 
 500 & 00 40 52.88 & +40 36 24.4 & 3.07   & 5.6e--3$\pm$6.0e--5 &  94 & 0.69$\pm$0.30 &   0.70$\pm$0.11 & --0.18$\pm$0.10 & --0.42$\pm$0.23 \\ 
 509 & 00 40 55.02 & +41 12 16.4 & 5.23   & 2.2e--3$\pm$4.0e--5 &  12 & 0.52$\pm$0.19 & --0.67$\pm$0.19 &                 &                 \\ 
 521 & 00 40 58.94 & +41 03 00.6 & 6.72   & 2.2e--3$\pm$5.0e--5 &  10 & 0.80$\pm$0.19 & --0.46$\pm$0.20 & --0.88$\pm$0.38 &                 \\ 
 560 & 00 41 08.08 & +40 31 42.4 & 4.04   & 5.6e--3$\pm$7.0e--5 &  53 & 0.66$\pm$0.11 & --0.71$\pm$0.12 & --0.16$\pm$0.40 &                 \\ 
 682 & 00 41 40.28 & +40 59 47.9 & 5.11   & 1.6e--3$\pm$4.0e--5 & 7.6 & 0.61$\pm$0.23 & --0.48$\pm$0.26 &   0.11$\pm$0.38 & --0.27$\pm$0.57 \\
 969 & 00 42 39.82 & +40 43 18.8 & 1.84   & 7.3e--2$\pm$2.0e--4 &4800 & 0.06$\pm$0.02 & --0.75$\pm$0.02 & --0.92$\pm$0.07 &                 \\
1079 & 00 42 55.50 & +40 59 46.4 & 3.00   & 5.1e--3$\pm$6.0e--5 & 110 & 0.77$\pm$0.09 & --0.69$\pm$0.11 & --0.90$\pm$0.37 &                 \\ 
1083 & 00 42 56.78 & +40 57 18.5 & 3.51   & 2.0e--3$\pm$3.0e--5 &  28 & 0.43$\pm$0.14 & --0.98$\pm$0.09 &                 &                 \\ 
1148 & 00 43 08.85 & +41 03 05.4 & 5.25   & 2.7e--3$\pm$5.0e--5 &  16 & 0.66$\pm$0.16 & --0.60$\pm$0.21 &                 &                 \\ 
1156 & 00 43 10.43 & +41 38 50.1 & 4.80   & 2.3e--3$\pm$4.0e--5 &  24 &               &   0.98$\pm$0.32 &   0.25$\pm$0.19 &   0.10$\pm$0.19 \\ 
1282 & 00 43 41.57 & +41 34 06.6 & 6.59   & 2.4e--3$\pm$3.0e--5 &  38 & 0.08$\pm$0.14 & --0.96$\pm$0.15 &                 &                 \\
1332 & 00 43 54.13 & +41 20 47.3 & 6.10   & 4.7e--3$\pm$8.0e--5 &  22 & 0.12$\pm$0.17 & --0.91$\pm$0.18 &                 &                 \\
1370 & 00 44 04.55 & +41 58 06.5 & 5.05   & 2.8e--3$\pm$6.0e--5 &  21 & 0.94$\pm$0.14 & --0.96$\pm$0.14 &                 &                 \\
1437 & 00 44 21.18 & +42 16 21.9 & 6.22   & 2.1e--3$\pm$6.0e--5 & 8.5 & 0.73$\pm$0.21 & --0.64$\pm$0.25 &                 &                 \\
1481 & 00 44 34.90 & +41 25 12.7 & 5.69   & 4.2e--3$\pm$8.0e--5 &  20 & 0.97$\pm$0.11 & --0.87$\pm$0.17 &                 &                 \\
1505 & 00 44 42.73 & +41 53 40.5 & 2.19   & 7.0e--3$\pm$5.0e--5 & 840 & 0.50$\pm$0.27 & --0.73$\pm$0.27 &   0.97$\pm$0.04 &   0.50$\pm$0.04 \\
1534 & 00 44 50.02 & +41 52 40.9 & 3.67   & 2.8e--3$\pm$5.0e--5 &  26 & 0.59$\pm$0.19 & --0.39$\pm$0.17 & --0.54$\pm$0.31 &                 \\
1548 & 00 44 55.73 & +41 56 55.2 & 4.24   & 1.9e--3$\pm$4.0e--5 &  20 & 0.95$\pm$0.13 & --0.34$\pm$0.19 & --0.83$\pm$0.34 &                 \\
1608 & 00 45 17.87 & +42 05 02.1 & 6.19   & 2.4e--3$\pm$5.0e--5 &  14 & 0.57$\pm$0.20 & --0.53$\pm$0.25 &                 &                 \\
1637 & 00 45 28.35 & +41 46 05.9 & 2.58   & 5.7e--3$\pm$5.0e--5 & 170 & 0.34$\pm$0.33 &   0.70$\pm$0.11 & --0.06$\pm$0.10 & --0.06$\pm$0.15 \\
1669 & 00 45 38.26 & +41 12 46.9 & 5.13   & 3.4e--3$\pm$1.0e--4 &  16 & 0.65$\pm$0.31 & --0.61$\pm$0.38 &                 &                 \\
1712 & 00 45 56.01 & +42 11 17.2 & 4.60   & 3.4e--3$\pm$7.0e--5 &  18 & 0.61$\pm$0.26 &   0.04$\pm$0.25 &   0.02$\pm$0.24 & --0.15$\pm$0.37 \\
1732 & 00 46 02.53 & +41 45 13.2 & 4.11   & 4.3e--3$\pm$7.0e--5 &  28 & 0.95$\pm$0.64 &   0.52$\pm$0.19 & --0.13$\pm$0.17 & --0.16$\pm$0.30 \\
1741 & 00 46 04.42 & +41 49 43.2 & 3.09   & 5.6e--3$\pm$7.0e--5 &  90 & 0.33$\pm$0.11 & --0.65$\pm$0.14 & --0.78$\pm$0.42 &                 \\
1748 & 00 46 06.34 & +41 29 23.6 & 4.21   & 2.4e--3$\pm$4.0e--5 &  27 & 0.58$\pm$0.19 & --0.40$\pm$0.21 & --0.17$\pm$0.35 &  0.54$\pm$0.23  \\
1796 & 00 46 25.39 & +41 09 38.7 & 5.06   & 9.7e--3$\pm$2.0e--4 &  32 & 0.33$\pm$0.22 & --0.48$\pm$0.21 &   0.33$\pm$0.27 & --0.59$\pm$0.43 \\
\hline                                    
\end{tabular}
\tablefoot{
\tablefoottext{a}{Detection likelihood.}
\tablefoottext{b}{Hardness ratios.}
\tablefoottext{c}{Classification by SPH11.}
}
\end{table*}

\end{appendix}

\clearpage

\begin{appendix}

\section{SDSS images}\label{sdss3}

{\bf See 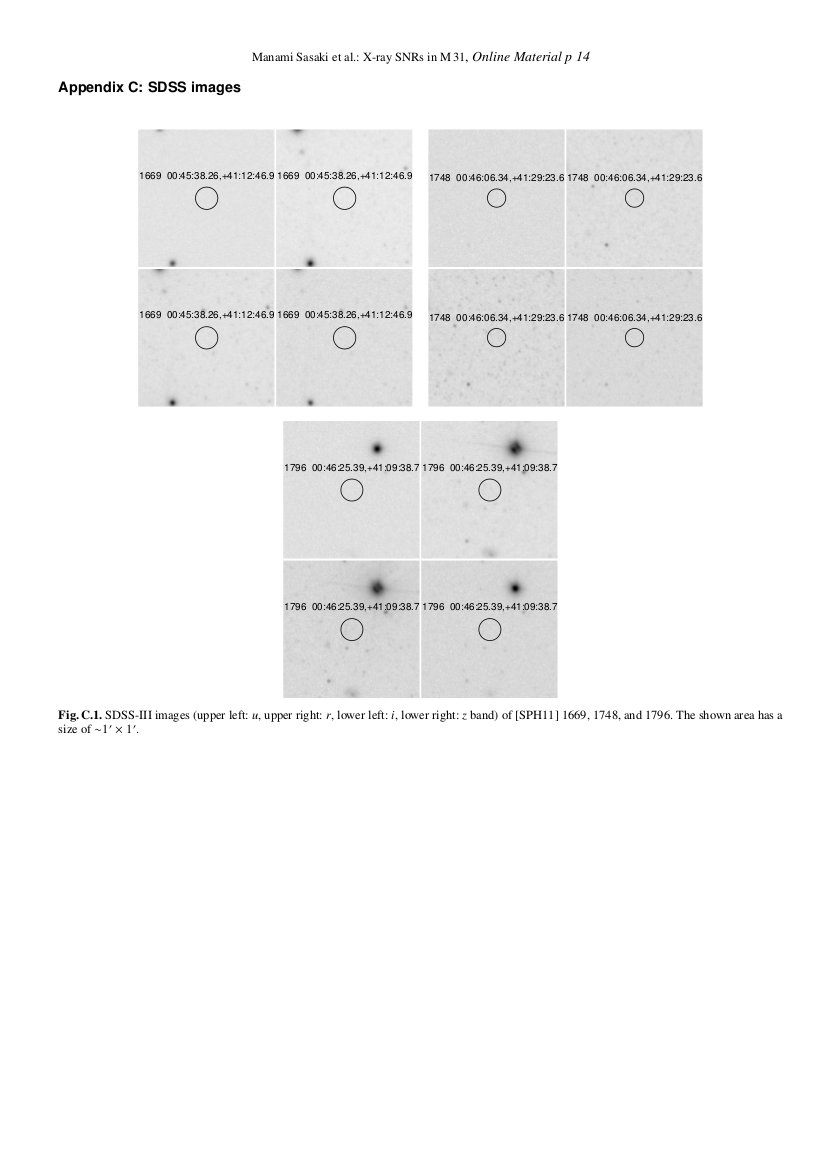.}

\end{appendix}

\clearpage

\begin{appendix}

\twocolumn

\section{Sources with unchanged classification}\label{unchanged}

\subsection{SNRs}

\subsubsection{[SPH11] 182: XMMM31~J003923.50+404419.1}

[SPH11] 182 is a soft X-ray source
with hardness ratios 
indicative of an SNR.
No foreground star was found at its position.
The X-ray source is coincident with the optical SNR [DDB80] 1-4,
which corresponds to [BA64] 474 and [PAV78] 118.
Based on these results the source was classified as an SNR by SPH11.

The optical source is clearly visible in all three emission line images
of the LGGS (Fig.\,\ref{lggs0}) and has a limb-brightened shell that is 
bright in the southeast. The diameter of the optical SNR is $\sim$16\arcsec.
We confirm the SNR classification based on 
the [\sii]/\ha\ flux ratio (Table \ref{snrlist}).
The LGGS $UBVRI$ band images show several blue point-like sources at the 
position of the SNR indicating a group of young stars. 

\subsubsection{[SPH11]  474: XMMM31~J004047.19+405524.6}

[SPH11] 474 has hardness ratios indicative of an SNR
and is coincident with the optical SNR [DDB80] 1-7. 
A radio counterpart [GLG04] 68 was found with $\alpha = -0.25$, which 
might indicate a PWN (GLG05). 

From the analysis of the LGGS narrow band images, we derive an
[\sii]/\ha\ flux ratio of $1.0\pm0.5$ (Fig.\,\ref{lggs}2) and thus
confirm the SNR identification. The source is rather
compact with a diameter of $\sim$5\arcsec, corresponding to 
$\sim$18~pc. 

\subsubsection{[SPH11]  883: XMMM31~J004224.41+411729.8}

[SPH11] 883 has hardness ratios indicative of an SNR
and is coincident with the \chandra\ source
[KGP02] r2-57, which has been identified as an SNR by WSK04 based on 
additional studies of LGGS and Very Large Array radio data. 

Its optical line flux ratio is [\sii]/\ha\ = $0.9\pm0.2$, corroborating
the SNR nature.
The optical source has an arc-like structure with an extent of $\sim$9\arcsec\
(Fig.\,\ref{lggs}4).
It is brightest in the east.
This source is the second closest SNR to the center of the galaxy with a 
galactocentric distance of about 860~pc.

\subsubsection{[SPH11] 1040: XMMM31~J004249.11+412406.6}

[SPH11] 1040 
is coincident with the \chandra\ source [KGP02] r3-84,
which has been identified as an SNR by WSK04, as well as
with the radio source [B90] 97. Therefore, it was classified as an SNR
by PFH05 and SPH11.

There is significant \ha, [\sii], and [\oiii] emission at its position
and the flux ratio of [\sii]/\ha\ = $0.8\pm0.2$ confirms its SNR nature 
(Fig.\,\ref{lggs}4).
The optical SNR has a patchy, ring-like morphology with an extent of 
$\sim$6\arcsec\ and is brightest in the north. 
It is surrounded by faint, diffuse \ha\ emission.

\subsubsection{[SPH11] 1050: XMMM31~J004250.47+411556.7}\label{1050}

[SPH11] 1050 is a soft source 
and is coincident with the \chandra\ source [KGP02] r2-56, which was identified 
as an SNR by \citet{2003ApJ...590L..21K}. Therefore, the \xmm\ source
was classified as an SNR by SPH11. The source also has an extended radio 
counterpart [B90] 101. It is located at a distance of about 
70\arcsec\ from the galactic center (i.e., about 250~pc). 

In the optical the SNR is a rather compact source with an extent of 
$\sim$3\arcsec\ (Fig.\,\ref{lggs}4). 
On the LGGS images, it is located at the edge of 
the central part of M\,31 where the subtracted continuum images are 
saturated as can be seen on the [\sii]/\ha\ ratio image.

\subsubsection{[SPH11] 1066: XMMM31~J004253.53+412551.0}\label{1066}

[SPH11] 1066 is a bright X-ray source with a detection likelihood of 
$ML = 1400$. 
It is coincident with a \chandra\ source that has been classified as 
an SNR by KGP02 (r3-69), the optical SNR [DDB80] 1-13 ([BA64] 521), and 
its radio counterpart [B90] 106. Therefore, the \xmm\ source was
classified as an SNR by PFH05 and SPH11. 

There is significant \ha, [\sii], and [\oiii] emission at its position
and the flux ratio of [\sii]/\ha\ = $0.9\pm0.2$ confirms its SNR nature 
(Fig.\,\ref{lggs}4).
It is slightly elongated in the north-south direction with a size
of $\sim$4\arcsec\ $\times$ 8\arcsec, being brighter in the south.
The relatively hard X-ray spectrum and the rather small extent of the source 
seen in the optical suggest that it is a young SNR.

\subsubsection{[SPH11] 1234: XMMM31~J004327.93+411830.5}\label{1234}

[SPH11] 1234 is the brightest source in our list with $ML = 6300$ and 
is a soft X-ray source.
It is coincident with the \chandra\ source [KGP02] r3-63 identified as
an SNR by \citet{2002ApJ...580L.125K},
the optical SNR [MPV95] 2-033, and the radio source [B90] 142. 
Therefore, it was classified as an SNR by PFH05 and SPH11.

Its \xmm\ spectrum is best fitted with a combination of two thermal 
components (see Sect.\,\ref{xrayspec}).
The soft X-ray spectrum indicates that it is not strongly 
absorbed. In addition, since there is
significant emission even below 0.5~keV but almost no emission above
1~keV this source might be a rather old SNR. The optical source
seen on the LGGS \ha, [\sii], and [\oiii] images has a horse-shoe shaped
structure with an extent of $\sim$8\arcsec, corresponding
to $50 - 60$~pc (see Fig.\,\ref{lggs}6). It is brighter on one side (northwest)
and is open to the other side (southeast), typical for SNRs evolving in a 
medium with a density gradient. 
There is also significant extended diffuse emission visible in the optical 
emission line images around the SNR,
which indicates that the SNR might be embedded in a superbubble. The X-ray 
spectrum of the SNR might therefore be contaminated by thermal emission 
from the interstellar gas in its environment.

\subsubsection{[SPH11] 1275: XMMM31~J004339.27+412653.0}

[SPH11] 1275 is a bright source ($ML = 1200$) embedded in a region with
soft diffuse emission. 
The X-ray position agrees very well with the optical SNR [MPV95] 3-059, 
which is also a radio source ([B90] 158). Therefore, the source was 
classified as an SNR by PFH05 and SPH11.

Its  [\sii]/\ha\ flux ratio is 0.9$\pm$0.2. The SNR has a half-circular shape
in the optical, which is open to the northwest and has an extent of 
$\sim$7\arcsec\ (Fig.\,\ref{lggs}6). 

\subsubsection{[SPH11] 1291: XMMM31~J004343.97+411232.7}

[SPH11] 1291 is a soft X-ray source 
and has been classified as an SNR 
based on its positional coincidence with the radio source
[B90] 166, which is an extended continuum source. 
O06 also discussed the possibility of this source
being an SNR ([O06] 4).

In the optical it is located in between two diffuse sources in the east 
and the south visible in \ha, most likely \hii\ regions (Fig.\,\ref{lggs}6). 
There is neither significant [\sii] nor [\oiii] emission. 
Therefore, the source is most likely a non-radiative SNR.

\subsubsection{[SPH11] 1328: XMMM31~J004353.69+411204.4}

[SPH11] 1328 is a soft X-ray source
coincident with the optical SNR [BW93] K230A
(Fig.\,\ref{lggs}7) and was thus classified as an SNR by PFH05
and SPH11. Radio emission that fills 
the western half of the gap between the optical line emission was
also found. 

In the optical, the source consists of brighter regions in the north and
the south, which are both extended to the east-west direction, thus 
forming a parallel structure. The entire structure has an extent of
$\sim$15\arcsec.

\subsubsection{[SPH11] 1351: XMMM31~J004358.25+411328.7}

[SPH11] 1351 is a soft source
coincident with the optical SNR [BW93] K252 
([DDB80] 18) and the radio source [B90] 199. 
Therefore, it was classified as an SNR by PFH05 and SPH11.

In the optical the source has an extent of $\sim$12\arcsec\ and is 
located on the southern edge of the more extended structure visible 
on the \ha\ image (Fig.\,\ref{lggs}7).
The [\sii]/\ha\ flux ratio of $0.8\pm0.2$ fulfills the criterion for SNRs.
In the optical, the source has a half-shell structure in the south with
filaments extending to the north. In general, it has a patchy morphology.

\subsubsection{[SPH11] 1386: XMMM31~J004409.53+413320.9}

[SPH11] 1386 is a soft X-ray source.
In the catalogues of PFH05 and SPH11 it was
classified as an SNR based on the positional coincidence with
the radio source [B90] 217.

It is located inside the \hii\ region [WB92] 315,
which can also be seen in the LGGS images (Fig.\,\ref{lggs}8). 
The [\sii]/\ha\ flux ratio at the X-ray position is low ($\approx0.2$). 
Therefore, the source is likely a non-radiative SNR.

\subsubsection{[SPH11] 1410: XMMM31~J004413.55+411954.3}

[SPH11] 1410 is a soft source
coincident with the optical SNR 
[BW93] K327 with the radio counterpart [B90] 224. Therefore, it was
classified as an SNR by SPH11.

The X-ray position coincides with a knot with an extent of
$\sim$4\arcsec\ in the optical emission line
images located at the southern edge 
of a more extended diffuse emission (Fig.\,\ref{lggs}8). 
The extended emission is likely an
\hii\ region, whereas the optical SNR has a flux ratio of 
[\sii]/\ha\ = $0.9\pm0.2$. 

\subsubsection{[SPH11] 1497: XMMM31~J004438.91+412528.8}

[SPH11] 1497 is a faint soft source 
and has been classified as an SNR by PFH05 and SPH11 based on its
coincidence with an SNR candidate suggested by [MPV95]. It is the 
counterpart of the optical and radio SNR [BW93] K506A, [B90] 278. 

Its morphology of the optical line emission is that of a circular shell 
with a diameter of $\sim$12\arcsec\ and is best seen 
in the [\sii] image (Fig.\,\ref{lggs}9). There is a bright blob-like 
\ha\ source east to the SNR. While the SNR shell has an [\sii]/\ha\ flux 
ratio of $\sim$1.0, the \ha\ blob makes the [\sii]/\ha\ ratio of the
total SNR lower ($0.5\pm0.1$).

\subsubsection{[SPH11] 1522: XMMM31~J004447.19+412918.7}

[SPH11] 1522 is a soft X-ray source 
and coincides with the optical SNR [BW93] K548
and was thus identified as an SNR by PFH05 and SPH11.

A shell with an extent of $\sim$10\arcsec\ is seen on the LGGS images, 
which is slightly elongated in the
east-west direction (Fig.\,\ref{lggs}9) and is brighter to the west.
Its [\sii]/\ha\ flux ratio of $1.0\pm0.2$ is clearly indicative of an SNR.
A nebula bright only in \ha\ with a similar size is located northeast
of the SNR.

\subsubsection{[SPH11] 1539: XMMM31~J004452.82+415458.1}

[SPH11]1539 
is the X-ray counterpart of the optical SNR [BW93] K594 and 
its radio counterpart [B90] 316. Therefore, it was identified as
an SNR by PFH05 and SPH11.

It has an extent of $\sim$8\arcsec\ in the optical and is brighter in the 
north (Fig.\,\ref{lggs}10).
Its [\sii]/\ha\ flux ratio is 1.0$\pm$0.4.

\subsubsection{[SPH11] 1587: XMMM31~J004510.59+413251.3}

[SPH11] 1587 is a faint, soft X-ray source
and is positionally coincident with a shell-type radio source [B90] 354
with an offset of $\sim$3\arcsec. It has thus been classified as an SNR 
by SPH11.

No optical emission is detected at its position. Therefore, 
the SNR seems to be non-radiative.

\subsubsection{[SPH11] 1593: XMMM31~J004512.31+420029.6}

[SPH11] 1593 is a faint, soft X-ray source 
and is coincident with a radio source [B90] 365 with an amorphous shape
and was thus classified as an SNR by SPH11.

Faint shell-like emission can be seen in \ha, [\sii], [\oiii] images
(Fig.\,\ref{lggs}10) with a diameter of about 30\arcsec. Its
optical flux ratio [\sii]/\ha\ is 1.0$\pm$0.4, indicating its SNR nature.
Therefore, we also classify this source as an SNR, most likely a highly evolved
SNR. It is so far the largest X-ray SNR detected in M\,31.

\subsubsection{[SPH11] 1599: XMMM31~J004513.94+413615.5}

[SPH11] 1599 is the X-ray counterpart 
of the optical SNR [DDB80] 19 ([BW93] K717) and the radio source [B90] 
367. It was identified as an SNR by PFH05 and SPH11.

The optical SNR has an extent of $\sim$10\arcsec. It is bright in the west 
and has a diffuse extension to the east (Fig.\,\ref{lggs}10). 
The X-ray poisition coincides with the bright part in the optical.

\subsubsection{[SPH11] 1793: XMMM31~J004624.71+415541.6}

[SPH11] 1793 is a soft X-ray source
coincident with the source [PFH05] 745. The X-ray source has 
an extended radio counterpart [B90] 472 and was thus
classified as an SNR by PFH05 and SPH11.

Only faint emission is seen on the LGGS images (Fig.\,\ref{lggs}12). 
The optical flux ratio [\sii]/\ha\ is 0.7$\pm$0.3.
Based on the X-ray properties and the coincidence with
a radio continuum source, we also classify this source as
an SNR, which is most likely non-radiative.
 
\subsubsection{[SPH11] 1805: XMMM31~J004627.91+420806.7}

[SPH11] 1805 is a soft X-ray source 
coincident with the radio source [B90] 476
and was thus classified as an SNR by PFH05, O06, and SPH11.

Very faint optical line emission is detected on the LGGS images.
The [\sii]/\ha\ flux ratio of 3.0$\pm$1.1 is indicative
of an SNR.

\subsection{SNR candidates}

\subsubsection{[SPH11]  294: XMMM31~J003958.28+402726.2}

[SPH11] 294 is a bright X-ray source with 
hardness ratios indicative of an SNR. As no foreground star is
detected it has been suggested as an SNR candidate by 
PFH05 and SPH11.

We have determined \ha\ and [\sii] fluxes from the LGGS images, 
however, the source is very faint (Fig.\,\ref{lggs}1). 
Therefore, we keep the classification of this source as an SNR 
candidate. If it is an SNR, the optical fluxes indicate that it
is likely non-radiative.

\subsubsection{[SPH11]  414: XMMM31~J004030.46+402756.0}

[SPH11] 414 is a hard source and thus was not classified as
an SNR candidate based on the hardness ratios
but has been suggested as an SNR candidate based on the positional 
coincidence with the radio source [GLG04] 198 with a radio spectral 
index of $\alpha = -0.08$. Its radio properties indicate 
emission from a PWN (GLG05).

We detect neither \ha, [\sii], nor [\oiii] emission on the LGGS 
images (Fig.\,\ref{lggs}1). Therefore, if the source is an SNR, it 
has to be non-radiative. A PWN might produce hard X-ray emission. 
Based on the radio coincidence, we keep its identification as an 
SNR candidate.

\subsubsection{[SPH11]  419: XMMM31~J004031.92+405837.4}

[SPH11] 419 is a soft X-ray source with hardness ratios 
indicative of an SNR.
No optical source is detected at its position that might
be a foreground star. Therefore, it has been classified as an SNR 
candidate by SPH11. 

We do not detect any emission in the narrow-band filter images of 
the LGGS (Fig.\,\ref{lggs}1). If the source is an SNR it has to be 
non-radiative. Because of lack of additional information, we keep the SNR 
candidate classification. 

\subsubsection{[SPH11]  441: XMMM31~J004040.39+403012.3}

[SPH11] 441 is a soft X-ray source with hardness ratios 
indicative of an SNR. No foreground star was found
at its position. Therefore, it has been classified as an SNR
candidate by PFH05 and SPH11.

Faint \ha\ and [\sii] emission was detected in the LGGS images 
(see Table \ref{candlist} and Fig.\,\ref{lggs}1) on the 
northwestern side.
However, this emission is not significant to classify the
optical source as an SNR. Therefore, the \xmm\ source remains an 
SNR candidate.

\subsubsection{[SPH11]  521: XMMM31~J004058.94+410300.6}

[SPH11] 521 is one of the faintest X-ray sources with a detection 
likelihood of $ML = 10$. Its hardness ratios 
suggest that it is an SNR candidate. No optical counterpart was found
that could be a foreground star. It was therefore proposed
as an SNR candidate by PFH05 and SPH11.

The X-ray source seems to be surrounded by a very extended optical source, 
which is visible in \ha, [\sii], and [\oiii]  (Fig.\,\ref{lggs}2) and 
has a low flux ratio [\sii]/\ha\ = $0.3\pm0.2$, hence is an \hii\ region.
The X-ray source therefore remains an SNR candidate.

\subsubsection{[SPH11]  560: XMMM31~J004108.08+403142.4}

[SPH11] 560 is a soft X-ray source,
which was classified as an SNR candidate by PFH05 and SPH11 
based on its hardness ratios and the lack of a foreground star as a 
possible optical counterpart.

Faint \ha\ and [\sii] emission was measured at its position on the LGGS images.
This source thus remains an SNR candidate.

\subsubsection{[SPH11]  969: XMMM31~J004239.82+404318.8}\label{969}

[SPH11] 969 is a bright X-ray source with a detection likelihood of
$ML = 4800$. Its hardness ratios
suggest that it is an SNR candidate. As no foreground star was found,
the \xmm\ source was suggested to be an SNR candidate by SPH11.
The X-ray source has also been detected with \chandra\ ([WGK04] s1-84).

There is neither a point source nor significant extended emission in the 
optical at and around the source. 
Without any confirmation in optical or radio, this source is still classified 
as an SNR candidate.

\subsubsection{[SPH11] 1083: XMMM31~J004256.78+405718.5}

[SPH11] 1083 is a soft X-ray source 
with no foreground star at its position and hence has been classified as
an SNR candidate by SPH11.

No optical source is detected at its position on the LGGS images.
Therefore, if it is an SNR, it is non-radiative.
The source remains an SNR candidate.

\subsubsection{[SPH11] 1282: XMMM31~J004341.57+413406.6}

[SPH11] 1282 is a soft X-ray source with no foreground star as a
possible counterpart
and thus was classified
as an SNR candidate by SPH11. PFH05 classified the X-ray as a candidate for a
super-soft source. 

There is faint optical line emission on the LGGS image 
with an extent of $\sim$14\arcsec\ (Fig.\,\ref{lggs}6). 
The flux ratio of $0.5\pm0.1$
is not significantly high enough to classify the optical source as an SNR.
It is very diffuse and patchy, similar to [SPH11] 1148 and 1328.
This source still remains an SNR candidate based on its X-ray hardness ratios.

\subsubsection{[SPH11] 1332: XMMM31~J004354.13+412047.3}

[SPH11] 1332 is a soft X-ray source with no foreground star
as possible counterpart,
which was classified as an SNR candidate by PFH05 and SPH11 based
on its hardness ratios.

There is faint \ha\ and [\oiii] emission confirmed in the LGGS data, 
however no [\sii] emission was detected (see Table \ref{candlist} 
and Fig.\,\ref{lggs}7).
A filament in \ha\ is visible with an extent of $\sim$5\arcsec.
This source remains an SNR candidate.

\subsubsection{[SPH11] 1669: XMMM31~J004538.26+411246.9}

[SPH11] 1669 is a faint soft X-ray source,
which was classified as an SNR candidate by SPH11 
based on its hardness ratios and the lack of a foreground star as
its possible counterpart.

The position of the X-ray source was not observed by the LGGS. No optical 
counterpart is found on the SDSS-III images 
(Fig.\,\ref{sdss3}, upper left).
Radio emission has not been detected at its position so far either. 
This source remains an SNR candidate based on its X-ray properties.

\subsubsection{[SPH11] 1748: XMMM31~J004606.34+412923.6}

[SPH11] 1748 is a faint soft X-ray source, which marginally
fulfilled the hardness ratio criterion for SNR candidates
and has no optical counterpart indicative of a star.
Therefore, it was classified as an SNR candidate by SPH11.

The LGGS did not cover the position of the X-ray source.
The SDSS-III images show no source at the \xmm\ position inside 
the error circle 
(Fig.\,\ref{sdss3}, upper right).
A radio counterpart is not known either. 
Therefore, this source remains an SNR candidate.

\subsubsection{[SPH11] 1796: XMMM31~J004625.39+410938.7}

[SPH11] 1796 
was classified as an SNR candidate by PFH05 and SPH11 based on $HR_1$
and $HR_2$ and the lack of a foreground star at its position.

The source was not covered by the LGGS. The SDSS-III images show no 
source inside the \xmm\ positional error circle 
(Fig.\,\ref{sdss3}, lower panel). 
A radio counterpart is not known either. 
With no additional information available, this source remains
an SNR candidate.

\subsection{No SNRs}

\subsubsection{[SPH11] 1121: XMMM31~J004303.70+413717.2}

[SPH11] 1121 is a hard source
and was not suggested to be an SNR candidate by SPH11.
However, we have studied this source because of a coincident diffuse
optical emission seen in \ha\ and [\sii].

Diffuse emission is seen both on the \ha\ and [\sii] LGGS images 
(Fig.\,\ref{lggs}5). However, the flux ratio is
[\sii]/\ha\ = $0.2\pm0.1$, not indicative of an SNR. The diffuse optical source
at and north of the X-ray emission is most likely an \hii\ region. 
As this source is hard in X-rays and there is no optical counterpart indicative
of an SNR, we do not classify it as an SNR candidate and keep the
$<$hard$>$ classification.

\subsubsection{[SPH11] 1461: XMMM31~J004428.62+414948.7}

[SPH11] 1461 is a soft source 
and was classified as a foreground star candidate by SPH11 because
stars with characteristic \logfxfo\ values are found in the 
X-ray error circle, with an optically red object close to the
center of the X-ray error circle. However, the \ha\ image also
revealed a diffuse source right next to the X-ray source in the north.

The diffuse optical source is visible in \ha, [\sii], and [\oiii] emission 
(Fig.\,\ref{lggs}8).
The flux ratio [\sii]/\ha\ = $0.2\pm0.1$, indicative of an \hii\ region.
Therefore, the origin of the soft X-ray emission seems to be either a star 
or an \hii\ region and the source is no SNR candidate.

\subsubsection{[SPH11] 1468: XMMM31~J004430.56+412306.2}

[SPH11] 1468 is a hard X-ray source.
As diffuse \ha\ emission was found in the optical
we considered this source for further studies.

The LGGS images show an extended diffuse source in \ha\ and [\sii] 
(Fig.\,\ref{lggs}8).
The [\sii]/\ha\ flux ratio is $0.4\pm0.1$. This value is at the border between
SNRs and \hii\ regions. While the more extended optical source seems to be an
\hii\ region, we cannot rule out that there is an SNR located at the same position.
However, as the X-ray emission does not indicate an SNR, we do not classify 
[SPH11] 1468 as an SNR candidate.

\subsubsection{[SPH11] 1611: XMMM31~J004518.44+413936.3}

[SPH11] 1611 is a hard X-ray source 
coincident with [PFH2005] 626 and [SBK2009] 220 and was classified
as a hard source by SPH11.
Diffuse optical \ha\, [\sii], and [\oiii] emission was found
around the X-ray source.

The extended optical source can be seen on the LGGS emission line
images (Fig.\,\ref{lggs}11). The X-ray source is located close to the 
northern rim of the optical source, coincident with a brighter 
shell-like structure. The entire optical source has a flux ratio of
[\sii]/\ha\ = 0.4 -- 0.5. 
The optical source is most likely a large \hii\ region.
The X-ray hardness ratios do not even marginally indicate an SNR.
Therefore, we do not consider this sources as an SNR candidate.

\end{appendix}

\end{document}